\documentclass[onecolumn,showpacs,preprintnumbers]{revtex4}
\usepackage{graphicx}
\usepackage{dcolumn}
\usepackage{bm}
\usepackage[english]{babel}

\begin{document}

\title{Nonlocal free energy of a spatially inhomogeneous superconductor.}
\author{K.V. Grigorishin}
\email{konst_dark@mail.ru}
\author{B.I. Lev}
\email{bohdan.lev@gmail.com} \affiliation{Boholyubov Institute for
Theoretical Physics of the Ukrainian National Academy of Sciences,
14-b Metrolohichna str. Kiev-03680, Ukraine.}
\date{\today}

\begin{abstract}

The microscopic approach was developed for obtaining of the free
energy of a superconductor based on direct calculation of the
vacuum amplitude. The free energy functional of the spatially
inhomogeneous superconductor in a magnetic field was obtained with
help of the developed approach. The obtained functional is
generalization of Ginzburg-Landau functionals for any temperature,
for arbitrary spatial variations of the order parameter and for
the nonlocality of a magnetic response and the order parameter.
Moreover, the nonlocality of the magnetic response is the
consequence of order parameter's nonlocality. The extremals of
this functional are considered in the explicit form in the low-
and high-temperature limit at the condition of slowness of spatial
variations of the order parameter.

\end{abstract}

 \maketitle

\section{Introduction}\label{intr}

One of the important applications of condensed matter physics is
the description of thermodynamics and electrodynamics of
superconductors. Several types of superconducting materials are
spatially inhomogeneous anisotropic structures. In particular,
heterostructures and superlattice, combinations of superconductors
with the materials with different types of a conductivity. The
superconductive and transport properties of the inhomogeneous
systems is determined by the proximity effect, which results to
modulation of the order parameter, and by Josephson effect
\cite{gog1,gog2}. At the present moment many various
heterostructures have been made on the basic of the high
temperature superconductors (HTS). However the HTS can be
considered as the heterostructures \citep{kapaev}, where the
conducting $CuO_{2}$ layers play a role quantum wells, and the
layers of atoms between them (the reservoirs of charge carrier)
play a role quantum barriers. The value of critical temperature
can essentially depend on the spatial structure of the
superconducting system. So, in the papers
\citep{ovch1,ovch2,ovch3} the model of cluster crystal (the
lattice is formed by metallic clusters) with the giant
intensification of a superconductive pairing (hypothetically the
critical temperature is $T_{C}\geq 150 K$) was proposed. Such
crystal can be considered as a superconducting superlattice. The
intensification of Cooper pairing can occur in a neighborhood of
the special defects in a solid \citep{khlyus}.

The modification of Ginzburg-Landau equation is necessary for
description of spatially inhomogeneous superconducting systems.
So, the modified functional of free anergy is proposed for the
description of influence of the defects \citep{khlyus}, where the
term $-\gamma\delta(x)|\Delta(x)|^{2}$ is introduced ($\gamma$ is
coupling constant, $\delta(x)$ - Dirac function, $|\Delta(x)|$ is
the gap function - order parameter). This term considers the
intensification of Cooper pairing around the defect, moreover the
area of intensification is much less than the correlation length
$l_{0}$ of the superconductor. For description of layered
compounds of type HTS and $MgB_{2}$ \citep{asker} the modified
Eliashberg equations with consideration of anisotropy and
Lawrence-Doniach functional \citep{lawr} for quasi two dimensional
superconductors are applied. The nonlocality causes additional
complication in the description of the spatially inhomogeneous
systems. The superconducting systems with a deep modulation of the
order parameter must be considered as the systems of Josephson
contacts. For the structures where $d\gg\Lambda$ ($d$ is the
thickness of a contact, $\Lambda$ is the magnetic penetration
depth) the nonlocality is the determining factor of
electrodynamics effects \citep{ivan1,ivan2,mints}. However the
effects of nonlocality can be essential even for the cases when
$d\ll\Lambda$ \citep{gurev,silin1,silin2}.

For the complete description of a superconductor it is necessary
to know the functional of free energy
$\Omega(\beta,\Delta,\textbf{A})$, where $\beta=\frac{1}{T}$ is
the inverse temperature, $\Delta$ is the gap function (order
parameter), $\textbf{A}$ is the potential of microscopic magnetic
field. Then the equations $\frac{\delta\Omega}{\delta\Delta}=0$,
$\frac{\delta\Omega}{\delta\textbf{A}}=0$ are the equations of
state of a superconductor. The basic method for obtaining of the
equations of state is solution of Gor'kov equations or Bogoliubov
equations with the equation of self-consistency for the order
parameter \citep{degen} $\Delta(\textbf{r})$ and
$\Delta^{+}(\textbf{r})$. However this set of equations is
nonlinear. If the temperature is close to the critical temperature
$T\rightarrow T_{C}\Rightarrow\Delta\rightarrow 0$, then these
equations can be represented in the form of series in degrees of
$\Delta$. Moreover, magnetic penetration depth $\Lambda$ must be
larger then Pippard coherent length $l_{0}$, hence the potential
$\textbf{A}$ changes a little on the coherent length. As a result
we have the well-known Ginzburg-Landau equations. The equations
are correct for description of thermodynamics and electrodynamics
of a superconductor under the following restrictions:
\begin{enumerate}
    \item The energy gap is much less than the critical temperature. Then the parameter $\Delta(\textbf{r},T)/T_{C}\ll
    1$ can be expansion parameter. This means, that the equations are correct in the range $T\rightarrow T_{C}$
    or in the range $H\rightarrow H_{C2}$ (intensity of magnetic field is close to the second critical magnetic field $H_{C2}$).
    \item $\Delta(\textbf{r},T)$ changes slowly on the coherent length $l(T)$, which is size
    of a Cooper pair.
    \item Magnetic field $\textbf{H}(\textbf{r})=\texttt{rot}\textbf{A}(r)$
     changes slowly on the coherent length, that is the magnetic penetration depth is $\Lambda(T)\gg l(0)$.
    This means, that electrodynamics of a superconductor is local.
\end{enumerate}

In the papers \citep{tew,werth} the equations have been proposed,
where the first restriction is absent. These equations were
obtained from Gor'kov equations and they are the generalization of
Ginzburg-Landau equations for the case of arbitrary value of
$\Delta(\textbf{r},T)/T$. However spatial inhomogeneities are
slow, the order parameter and magnetic response are local. In the
Ref. \citep{hook} on basic Bogoliubov equation it was shown that
if the gap $\Delta(\textbf{r})$ varies with position, that the
value of the gap in a point is determined by a distribution of gap
in some neighborhood: $\Delta(\textbf{r})=\int
d\textbf{r}'Q(\textbf{r},\textbf{r}')\Delta'(\textbf{r})$,
moreover the kernel $Q$ can be function of the order parameter
$\Delta(\textbf{r}')$ also. This means that the order parameter is
nonlocal.

The problem of description of the superconducting phase can be
solved by another method. This method is the direct calculation of
free energy. To this end we must calculate the partition function
$Z$
\begin{equation}\label{0.1}
 Z=Sp\left(Z_{0}\widehat{\rho}_{0}\widetilde{U}(\beta)\right)=Z_{0}R(\beta),
\end{equation}
where $Z_{0}$ is the partition function for a system of
noninteracting particles,
$\widetilde{U}(\beta)=e^{+\beta(\widehat{H}_{0}-\mu
\widehat{N})}e^{-\beta(\widehat{H}-\mu \widehat{N})}$ is the
evolution operator in the interaction representation (it describes
evolution of the system in imaginary time $it\rightarrow\tau$,
$\tau\in\left[0,\beta\right]$),
$\widehat{H}=\widehat{H}_{0}+\widehat{H}_{I}$ is the system
Hamiltonian, $\widehat{N}$ is the particle operator, $\mu$ is the
chemical potential (the replacement of Hamiltonian $\widehat{H}$
in canonical ensemble by Hamiltonian $\widehat{H}-\mu \widehat{N}$
in grand canonical ensemble leads to the shift of reference of
particle's energy from zero to Fermi surface:
$\varepsilon(k_{F})=0$),
\begin{eqnarray}\label{0.2}
R(\beta)=\langle
\widehat{U}(\beta)\rangle_{0}=Sp\left(\widehat{\rho}_{0}\widetilde{U}(\beta)\right)=\sum_{n=0}^{\infty}\frac{(-1)^{n}}{n!}\int_{0}^{\beta}d\tau_{1}\ldots\int_{0}^{\beta}d\tau_{n}
Sp\left(\widehat{\rho}_{0}
\widehat{T}\left\{\widehat{H}_{1}(\tau_{I})\ldots\widehat{H}_{I}(\tau_{n})\right\}\right)
\end{eqnarray}
is the vacuum amplitude of the system, where
$\widehat{H}_{I}(\beta)=e^{+\beta(\hat{H_{0}}-\mu
\hat{N})}\widehat{H}_{I}e^{-\beta(\hat{H_{0}}-\mu \hat{N})}$ is
the interaction operator of particles in interaction
representation, $\widehat{T}$ is the ordering operator in time.
The averaging $\langle\rangle_{0}\equiv
Sp(\widehat{\rho}_{0}\ldots)$ is done over ensemble of
\emph{noninteracting} particles.

The partition function $Z_{0}$ can be found exactly for any
system. If particles interact, then the situation becomes
essentially complicated. Solution of the problem reduces to the
calculation of a transition amplitude "vacuum-vacuum" (\ref{0.2}).
The vacuum amplitude determines the internal energy of a system:
\begin{equation}\label{0.3}
    U=-\frac{\partial}{\partial\beta}\ln Z_{0}-\frac{\partial}{\partial\beta}\ln
    R(\beta),
\end{equation}
and the grand thermodynamical potential:
\begin{equation}\label{0.4}
    \Omega=-\frac{1}{\beta}\ln Z_{0}-\frac{1}{\beta}\ln
    R(\beta).
\end{equation}
The potential $\Omega$ plays a role of Helmholtz free energy if a
particle's energy is counted from Fermi surface. Therefore we will
call the grand thermodynamical potential as free energy for
brevity.

Eqs. (\ref{0.3},\ref{0.4}) are valid if the symmetries of ground
state of the system with interaction $|\Psi_{0}\rangle$ and
without interaction $|\Phi_{0}\rangle$ are identical:
$\langle\Phi_{0}|\Psi_{0}\rangle\neq0$. This means that: 1)the
interaction potential is being switched slowly in the system in
the ground state without interaction, 2) the ground state of
system with interaction is being obtained by continuous way from
the ground state without interaction while the switching of the
interaction (\textit{adiabatic hypothesis}) \citep{matt1,matt2}.
If the wave functions are orthogonal
$\langle\Phi_{0}|\Psi_{0}\rangle=0$ - the adiabatic hypothesis is
not valid, and the symmetries of the system with interaction and
without one are different. This means, that the initial system
without interaction suffers a phase transition stipulated by the
interaction. For the system with the broken symmetry we can
calculate the vacuum amplitude on the free propagators
$G_{0}(\textbf{k},\tau)$ and use Eqs. (\ref{0.3},\ref{0.4}).
However we will find the wrong thermodynamics potentials. This
means that the state exists with more low energy than the obtained
value.

The method for direct calculation of vacuum amplitude $R(\beta)$
has been proposed in Ref.\citep{sad}. The concept lies in the fact
that we consider electrons in a \emph{normal} metal propagating in
a random "field" of fluctuations of the order parameter
$\Delta_{\textbf{q}}$, where $\textbf{q}$ is small wave-vector.
The operator of the interaction of electrons with the fluctuations
can be written as:
\begin{equation}\label{0.5}
    \widehat{H}_{int}=\sum_{\textbf{p}}\left[\Delta_{\textbf{q}}\widehat{C}_{\textbf{p}_{+}}^{+}\widehat{C}_{-\textbf{p}_{-}}^{+}+
    \Delta_{\textbf{q}}^{\ast}\widehat{C}_{-\textbf{p}_{-}}\widehat{C}_{\textbf{p}_{+}}\right],
\end{equation}
where $\textbf{p}_{\pm}=\textbf{p}\pm \textbf{q}/2$. The
correction to the thermodynamics potential is represented via the
vacuum amplitude $R$ utilizing Wick theorem:
\begin{eqnarray}\label{0.6}
    \Delta\Omega&=&-T\ln R(\beta)\approx-T[R(\beta)-1]=
    -\frac{T}{2!}\int_{0}^{1/T}d\tau_{1}\int_{0}^{1/T}d\tau_{2}\langle
    \widehat{T}_{\tau}(\widehat{H}_{int}(\tau_{1})\widehat{H}_{int}(\tau_{2}))\rangle+\ldots\nonumber\\
    &=&-T\int_{0}^{1/T}d\tau_{1}\int_{0}^{1/T}d\tau_{2}|\Delta_{\textbf{q}}|^{2}
    \sum_{\textbf{p}}G_{0}(\textbf{p}_{+},\tau_{1}-\tau_{2})G_{0}(-\textbf{p}_{-},\tau_{1}-\tau_{2})+\ldots.
\end{eqnarray}
The correction $\Delta\Omega$ is represented via the free
propagators $G_{0}$ of normal state only - we consider normal
metal at $T>T_{C}$, where the fluctuation sourse of Cooper pair
(\ref{1.27}) acts. As a result we have Landau expansion:
\begin{eqnarray}\label{0.7}
    \Omega_{s}-\Omega_{n}=\sum_{\textbf{q}}\left[\alpha(T)|\Delta_{\textbf{q}}|^{2}+\frac{b}{2}|\Delta_{\textbf{q}}|^{4}+\gamma q^{2}|\Delta_{\textbf{q}}|^{2}\right],
\end{eqnarray}
where $\alpha(T)\propto(T-T_{C}),b,\gamma$ are the expansion
coefficients. As for calculation of the correction $\Delta\Omega$
the free propagators $G_{0}$ of normal phase is used only, the
condensed phase is considered as fluctuations against the
background of the normal phase. This means, that the
high-temperature limit of superconductor's free energy functional
$\Omega (T\rightarrow T_{C})$ can be obtained by this method only.
Moreover, the hamiltonian (\ref{0.5}) does not contain any
parameters of a matter (for example, interaction constant between
fermions). In the present paper we proposed the method, which
permits to calculate the free energy functional of a spatial
inhomogeneous superconductor for any temperature. The vacuum
amplitude $R(\beta)$ is calculated on the interaction between
particles, that allows to use model potentials. This gives a
possibility to build the free energy functional of spatial
inhomogeneous superconductor in external fields proceeding from
first principles, unlike the method presented in \citep{sad} where
$R(\beta)$ is calculated on interaction of particles with a random
"field" of fluctuations of the order parameter
$\Delta_{\textbf{q}}$, and the critical temperature $T_{C}$ is
introduced by phenomenologically (its connection with the
interaction between particle is absent). Our method can be
generalized to second order phase transition in other systems,
because any phase transitions can be described in the formalism of
the anomalous Green function \citep{matt2}.

In the presented paper we have developed the microscopic approach
for finding of free energy functional of a superconductor
$\Omega(\Delta,T)$ with help of direct calculation of the vacuum
amplitude. This value is calculated on the dressed one-particle
propagators. The propagators are dressed due to the interaction of
free fermions with the fluctuations of pairing by the method
presented in  \citep{migdal,pines}. In the sections
\ref{spaceinhom} and \ref{magnetic} using the developed approach
of microscopic description of superconducting phase we obtain the
functional of free energy of a spatial inhomogeneous
superconductor in a magnetic field
$\Omega(\Delta_{\textbf{q}},\textbf{a}(\textbf{q}),T)$. This
functional generalizes Ginzburg-Landau functional for arbitrary
temperatures, for arbitrary spatial variations of the order
parameter and for nonlocality of the order parameter and a
magnetic response.

\section{The uncoupling of correlations method for calculation
of a vacuum amplitude.}\label{method}
\subsection{Normal and anomalous propagators.}

Let we have the system of $N$ noninteracting fermions in volume
$V$ at temperature $T$. For the description of this system we use
Matsubara propagators, where time $t$ is complex:
$t\rightarrow-i\tau$, $\tau\in[0,\beta]$. In ideal Fermy gas
propagation of a particle with momentum $\textbf{k}$, energy
$\varepsilon\approx v_{F}(|\textbf{k}|-k_{F})$ counted from Fermy
surface (we are using system of units, where $\hbar=k_{B}=1$) and
with spin $\sigma$ is described by the free propagator:
\begin{eqnarray}\label{1.1}
  &&G_{0}(\textbf{k},\tau=\tau_{2}-\tau_{1})=\left\{\begin{array}{cc}
    -i\texttt{Sp}\left\{\widehat{\rho}_{0}C_{\textbf{k},\sigma}(\tau_{2})C_{\textbf{k},\sigma}^{+}(\tau_{1})\right\}, \qquad \tau>0 \\
    i\texttt{Sp}\left\{\widehat{\rho}_{0}C_{\textbf{k},\sigma}^{+}(\tau_{1})C_{\textbf{k},\sigma}(\tau_{2})\right\}, \qquad \tau\leq0 \\
  \end{array}\right\}=\nonumber\\
&&-i\theta_{\tau}(g_{0}^{+}A_{0}e^{-|\varepsilon|\tau}+g_{0}^{-}B_{0}e^{|\varepsilon|\tau})+
i\theta_{-\tau}(g_{0}^{-}A_{0}e^{-|\varepsilon|\tau}+g_{0}^{+}B_{0}e^{|\varepsilon|\tau}),\qquad\theta_{\tau}=\left\{\begin{array}{c}
  1,\qquad \tau>0\\
  0,\qquad \tau<0\\
\end{array}\right\},\nonumber\\
&&G(\textbf{k},\tau)=\frac{1}{\beta}\sum_{n=-\infty}^{n=+\infty}G(\textbf{k},\omega_{n})e^{-i\omega_{n}\tau},
\qquad G(\textbf{k},\omega_{n})=\frac{1}{2}\int_{-\beta}^{\beta}G(\textbf{k},\tau)e^{i\omega_{n}\tau}d\tau,\qquad\omega_{n}=\frac{(2n+1)\pi}{\beta},\nonumber\\
&&G_{0}(\textbf{k},\omega_{n})=\frac{i}{i\omega_{n}-\varepsilon(k)}=i\frac{i\omega_{n}+\varepsilon}{(i\omega_{n})^{2}-\varepsilon^{2}}
=i\frac{A_{0}}{i\omega_{n}-|\varepsilon|}+i\frac{B_{0}}{i\omega_{n}+|\varepsilon|},
\end{eqnarray}
where
\begin{equation}\label{1.2}
A_{0}=\frac{1}{2}\left(1+\frac{\varepsilon}{|\varepsilon|}\right),\qquad
B_{0}=\frac{1}{2}\left(1-\frac{\varepsilon}{|\varepsilon|}\right),\qquad
g_{0}^{+}=\frac{1}{e^{-|\varepsilon|\beta}+1},\qquad
g_{0}^{-}=\frac{1}{e^{|\varepsilon|\beta}+1},
\end{equation}
$C_{\textbf{k},\sigma}(\tau)$ and
$C_{\textbf{k},\sigma}^{+}(\tau)$ are creation and annihilation
operators in Heisenberg representation, $\widehat{\rho}_{0}$ is
the density matrix of noninteracting particles:
\begin{equation}\label{1.3}
    \widehat{\rho}_{0}=\exp\left\{\frac{\Omega-\widehat{H}_{0}}{T}\right\}
=\exp\left\{\frac{\Omega-\sum_{\textbf{k},\sigma}\varepsilon(k)C_{\textbf{k},\sigma}^{+}C_{\textbf{k},\sigma}}{T}\right\}.
\end{equation}

Now let an attracting force acts between particles. Hamiltonian of
the system is
\begin{eqnarray}\label{1.4}
\widehat{H}_{0}+\widehat{H}_{I}&=&\sum_{\alpha}\sum_{\textbf{k}}\varepsilon\left(\textbf{k}\right)C_{\textbf{k},\alpha}^{+}C_{\textbf{k},\alpha}
+\frac{1}{2V}\sum_{\alpha,\gamma}\sum_{\textbf{k},\textbf{l}}V_{\textbf{l},-\textbf{l},\textbf{k},-\textbf{k}}g_{\alpha\gamma}C_{-\textbf{l},\gamma}^{+}
C_{\textbf{l},\alpha}^{+}C_{\textbf{k},\alpha}C_{-\textbf{k},\gamma}
\end{eqnarray}
The force is described by matrix element of interaction potential:
\begin{equation}\label{1.5}
    \langle
    \textbf{l},-\textbf{l}|\widehat{V}|\textbf{k},-\textbf{k}\rangle=\lambda
    w_{l}w_{k}<0, \qquad w_{k}=\left\{\begin{array}{c}
      1,\qquad \varepsilon(k)<\omega_{D} \\
      0,\qquad \varepsilon(k)>\omega_{D}\\
    \end{array}\right\},
\end{equation}
moreover interacting particles have opposite spins:
$g_{\alpha\gamma}=\left\{\begin{array}{c}
      1,\qquad \alpha\neq\gamma \\
      0,\qquad \alpha=\gamma\\
    \end{array}\right\}$.

\begin{figure}[h]
\includegraphics[width=8.5cm]{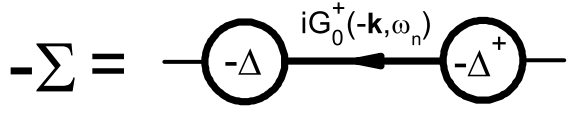}
\caption{The diagram for the mass operator  $\Sigma$ describing
interaction of a fermion with fluctuations of pairing.}
\label{fig1}
\end{figure}

The interaction (\ref{1.5}) leads to Cooper instability. Let the
additional particle with momentum $\{\textbf{k},\omega\}$
propagates through the system of identical fermions. A some pair
of fermions decays in components with momenta
$\{-\textbf{k},-\omega\}$ and $\{\textbf{k},\omega\}$ with
amplitude $i\Delta^{+}$. The second particle of the decayed pair
is in a state of the additional particle ($\textbf{k},\omega$) and
it is identical to the additional particle. The second particle
propagates through the system further. The first particle of
decayed pair forms bound state with the initial additional
particle with amplitude $-i\Delta$. Anew formed pair replenishes
the condensate of pairs in the system. Thus, the dressed
propagator $G_{S}$ considers the interaction of a particle with
the fluctuations of pairing. Intensity of the interaction is given
by the amplitudes $-i\Delta$ and $i\Delta^{+}$. Therefore we can
write the mass operator for such process (Fig.\ref{fig1}) as
\citep{migdal,pines}
\begin{equation}\label{1.6}
    -\Sigma(\textbf{k},\omega_{n})=(-\Delta)iG_{0}^{+}(-\textbf{k},\omega_{n})(-\Delta^{+})=\frac{-\Delta\Delta^{+}}{i\omega_{n}+\varepsilon(k)}.
\end{equation}

The dressed propagator can be found from Dyson equation:
\begin{eqnarray}\label{1.7}
&&\frac{1}{G_{0}}=\frac{1}{G_{S}}-i\Sigma\Rightarrow
G_{S}(\textbf{k},\omega_{n})=\frac{i}{i\omega_{n}-\varepsilon(k)-\Sigma(\textbf{k},\omega_{n})}\nonumber\\
&&=i\frac{i\omega_{n}+\varepsilon}{(i\omega_{n})^{2}-E^{2}(k)}=i\frac{A_{S}}{i\omega_{n}-E(k)}+i\frac{B_{S}}{i\omega_{n}+E(k)},
\end{eqnarray}
where $E=\sqrt{\varepsilon^{2}(k)+\Delta\Delta^{+}}$. It can be
written with help of a total definition of Green function in
$(\textbf{k},t)$-space:
\begin{eqnarray}\label{1.8}
  &&G_{S}(\textbf{k},\tau)=\tau_{2}-\tau_{1})=\left\{\begin{array}{cc}
    -i\texttt{Sp}\left\{\widehat{\varrho}_{0}C_{\textbf{k},\sigma}(\tau_{2})C_{\textbf{k},\sigma}^{+}(\tau_{1})\right\}, \qquad \tau>0 \\
    i\texttt{Sp}\left\{\widehat{\varrho}_{0}C_{\textbf{k},\sigma}^{+}(\tau_{1})C_{\textbf{k},\sigma}(\tau_{2})\right\}, \qquad \tau\leq0 \\
  \end{array}\right\}\nonumber\\
&&=-i\theta_{\tau}(g_{S}^{+}A_{S}e^{-E\tau}+g_{S}^{-}B_{S}e^{E\tau})+
i\theta_{-\tau}(g_{S}^{-}A_{S}e^{-E\tau}+g_{S}^{+}B_{S}e^{E\tau}),
\end{eqnarray}
where
\begin{equation}\label{1.9}
A_{S}=\frac{1}{2}\left(1+\frac{\varepsilon}{E}\right),\qquad
B_{S}=\frac{1}{2}\left(1-\frac{\varepsilon}{E}\right),\qquad
g_{S}^{+}=\frac{1}{e^{-E\beta}+1},\qquad
g_{S}^{-}=\frac{1}{e^{E\beta}+1}.
\end{equation}
$\widehat{\varrho}_{0}$ is the density matrix of noninteracting
\emph{quasi-particles}:
\begin{equation}\label{1.10}
    \widehat{\varrho}_{0}=\exp\left\{\frac{\Omega-\sum_{\textbf{k},\sigma}E(k)C_{\textbf{k},\sigma}^{+}C_{\textbf{k},\sigma}}{T}\right\}.
\end{equation}

Let us introduce the designations: $-G\Sigma\equiv\Delta F^{+},
-G^{+}\Sigma^{+}\equiv\Delta^{+}F$. Then Dyson equation can be
rewritten in a form of Gor'kov equations:
\begin{eqnarray}
  &&(i\omega_{n}-\varepsilon)G+\Delta F^{+}=i \label{1.11}\\
  &&(i\omega_{n}+\varepsilon)F^{+}+G\Delta=0\label{1.11a}.
\end{eqnarray}
The expressions for anomalous propagators follow from Gor'kov
equations:
\begin{equation}\label{1.12}
    F^{+}(\textbf{k},\omega_{n})=\frac{-i\Delta^{+}}{(i\omega_{n})^{2}-E^{2}(k)}, \qquad
F(\textbf{k},\omega_{n})=(F^{+}(\textbf{k},\omega_{n}))^{+}=\frac{i\Delta}{(i\omega_{n})^{2}-E^{2}(k)}.
\end{equation}
We can write the anomalous propagators in (\textbf{k},t)-space in
the form of vacuum averages of creation and annihilation
operators:
\begin{eqnarray}\label{1.13}
  F^{+}_{\alpha\gamma}(\textbf{k},\tau)&=&\frac{\Delta^{+}}{\sqrt{\Delta^{+}\Delta}}\left\{\begin{array}{cc}
    i\texttt{Sp}\{\widehat{\varrho}_{0}C_{-\textbf{k},\gamma}^{+}(\tau_{2})C_{\textbf{k},\alpha}^{+}(\tau_{1})\}, \qquad \tau>0 \\
    i\texttt{Sp}\{\widehat{\varrho}_{0}C_{\textbf{k},\alpha}^{+}(\tau_{1})C_{-\textbf{k},\gamma}^{+}(\tau_{2})\}, \qquad \tau\leq0 \\
  \end{array}\right\}\nonumber\\
&=&ig_{\alpha\gamma}\frac{\Delta^{+}}{\sqrt{\Delta^{+}\Delta}}\sqrt{A_{S}B_{S}}
\left[\left(g_{S}^{+}e^{-E\tau}-g_{S}^{-}e^{E\tau}\right)\theta_{\tau}-\left(g_{S}^{+}e^{E\tau}-g_{S}^{-}e^{-E\tau}\right)\theta_{-\tau}\right],\\
F_{\alpha\gamma}(\textbf{k},\tau)&=&\frac{\Delta}{\sqrt{\Delta^{+}\Delta}}\left\{\begin{array}{cc}
    -i\texttt{Sp}\{\widehat{\varrho}_{0}C_{\textbf{k},\alpha}(\tau_{2})C_{-\textbf{k},\gamma}(\tau_{1})\}, \qquad \tau>0 \\
    -i\texttt{Sp}\{\widehat{\varrho}_{0}C_{-\textbf{k},\gamma}(\tau_{1})C_{\textbf{k},\alpha}(\tau_{2})\}, \qquad \tau\leq0 \\
  \end{array}\right\}\nonumber\\
&=&ig_{\alpha\gamma}\frac{\Delta}{\sqrt{\Delta^{+}\Delta}}\sqrt{A_{S}B_{S}}
\left[-\left(g_{S}^{+}e^{-E\tau}-g_{S}^{-}e^{E\tau}\right)\theta_{\tau}+\left(g_{S}^{+}e^{E\tau}-g_{S}^{-}e^{-E\tau}\right)\theta_{-\tau}\right]
\label{5.13a}.
\end{eqnarray}
we can see, that the existence of nonzero anomalous propagators
$F$ and $F^{+}$ means breakdown of global gauge symmetry in a
system, that is the number of particles is not conserved in the
course of existence of a pair condensate.

\subsection{Kinetic energy and entropy.}

In order to calculate the free energy it is necessary to know the
kinetic energy of system particles, the energy of their
interaction and the entropy of collective excitations. The average
kinetic energy of all particles of a system is
\begin{eqnarray}\label{1.14}
    \langle
W\rangle&=&-2i\sum_{\textbf{k}}G(\textbf{k},\tau\rightarrow0^{-})\varepsilon(k)
=-\frac{2i}{\beta}\lim_{\tau\rightarrow0^{-}}\sum_{\textbf{k}}\sum_{n=-\infty}^{n=+\infty}G(\textbf{k},\omega_{n})e^{-i\omega_{n} t}\varepsilon(k)\nonumber\\
&=&2\sum_{\textbf{k}}(g^{-}A+g^{+}B)\varepsilon(k)=V2\nu_{F}\int_{-v_{F}k_{F}}^{\infty}(g^{-}A+g^{+}B)\varepsilon
d\varepsilon,
\end{eqnarray}
where $\nu_{F}=\frac{k_{F}^{2}}{2\pi^{2}v_{F}}$ is the density of
states on Fermy surface. Since the interaction (\ref{1.5}) exists
only in the layer $-\omega_{D}<\varepsilon(k)<\omega_{D}$, we can
suppose that
\begin{eqnarray}\label{1.15}
g^{-}=\left[
\begin{array}{cc}
  g_{0}^{-}; & |\varepsilon(k)|>\omega_{D} \\
  g_{S}^{-}; & |\varepsilon(k)|<\omega_{D} \\
\end{array}%
\right], \qquad g^{+}=\left[
\begin{array}{cc}
  g_{0}^{+}; & |\varepsilon(k)|>\omega_{D} \\
  g_{S}^{+}; & |\varepsilon(k)|<\omega_{D} \\
\end{array}%
\right].
\end{eqnarray}
Hence, one may write the expression for kinetic energy that
separats normal and superconductive parts as
\begin{eqnarray}\label{1.16}
\langle
W\rangle=W_{n}+V\nu_{F}\int_{-\omega_{D}}^{\omega_{D}}\tanh\left(\frac{\beta|\varepsilon|}{2}\right)\frac{\varepsilon^{2}}{|\varepsilon|}d\varepsilon-
V\nu_{F}\int_{-\omega_{D}}^{\omega_{D}}\tanh\left(\frac{\beta
E}{2}\right)\frac{\varepsilon^{2}}{E}d\varepsilon.
\end{eqnarray}
If we suppose that $\Delta=0$, then we shall have $W=W_{n}$.

At temperature $T\neq 0$ a gas of collective excitation exists -
boholons with the spectrum
$E=\sqrt{\varepsilon^{2}(k)+\Delta^{2}}$. Since boholons are
product of decay of Cooper pairs to fermions, hence the occupation
numbers of states by boholons are
\begin{equation}\label{1.17}
    f_{S}(k)=\frac{1}{e^{\beta E}+1}=\frac{1}{2}\left(1-\tanh\left(\frac{\beta
E}{2}\right)\right).
\end{equation}
Then the entropy of a system is
\begin{eqnarray}\label{1.18}
S&=&-2\sum_{\textbf{k}}\left[f(k)\ln f(k)+(1-f(k))\ln (1-f(k))\right]\nonumber\\
&=&S_{0}-2V\nu_{F}\int_{-\omega_{D}}^{\omega_{D}}\left[f_{S}\ln
f_{S}+(1-f_{S})\ln
(1-f_{S})\right]d\varepsilon+2V\nu_{F}\int_{-\omega_{D}}^{\omega_{D}}\left[f_{0}\ln
f_{0}+(1-f_{0})\ln (1-f_{0})\right]d\varepsilon.
\end{eqnarray}
Here we separated the normal part again, where
$f_{0}=(e^{\beta|\varepsilon|}+1)^{-1}$, so that $S=S_{n}$ at
$\Delta=0$. The multiplier "2" appeared as result of summation
over spin states.

\subsection{Vacuum amplitude.}

In the previous subsections we considered the interaction of
particles with fluctuations of pairing, and we found, that the
state of a system described by the density matrix
$\widehat{\varrho}_{0}$ has another symmetry in comparison with
the initial state $\widehat{\rho}_{0}$. The effective Hamiltonian
of the system of quasiparticle is
\begin{eqnarray}\label{1.19}
\widehat{H}_{0}+\widehat{H}_{I}&=&\sum_{\alpha}\sum_{\textbf{k}}E\left(\textbf{k}\right)C_{\textbf{k},\alpha}^{+}C_{\textbf{k},\alpha}
+\frac{1}{2V}\sum_{\alpha,\gamma}\sum_{\textbf{k},\textbf{l}}V_{\textbf{l},-\textbf{l},\textbf{k},-\textbf{k}}g_{\alpha\gamma}C_{-\textbf{l},\gamma}^{+}
C_{\textbf{l},\alpha}^{+}C_{\textbf{k},\alpha}C_{-\textbf{k},\gamma}
\end{eqnarray}
The vacuum amplitude of a system can be written in the form:
\begin{eqnarray}\label{1.20}
R(\beta)=\langle
\widehat{U}(\beta)\rangle_{0}=\texttt{Sp}\left(\widehat{\varrho}_{0}\widetilde{U}(\beta)\right)=\sum_{n=0}^{\infty}\frac{(-1)^{n}}{n!}\int_{0}^{\beta}d\tau_{1}\ldots\int_{0}^{\beta}d\tau_{n}
Sp\left(\widehat{\varrho}_{0}
T\left\{\widehat{H}_{1}(\tau_{I})\ldots\widehat{H}_{I}(\tau_{n})\right\}\right),
\end{eqnarray}
where
$\widehat{H}_{I}(\tau)=e^{+\tau\hat{H_{0}}}\widehat{H}_{I}e^{-\tau\hat{H_{0}}}$
is the interaction operator of particles in interaction
representation. The averaging
$\texttt{Sp}\left(\widehat{\varrho}_{0}\widetilde{U}(\beta)\right)$
is done over ensemble of \emph{noninteracting quasi-particles}. We
can write the expended expression for the vacuum amplitude:
\begin{eqnarray}\label{1.21}
&&R(\beta)=1+\frac{1}{1!}\frac{1}{V}\int_{0}^{\beta}d\tau_{1}\sum_{\alpha,\gamma}\sum_{\textbf{k},\textbf{l}}\left(-\frac{1}{2}V_{\textbf{l},-\textbf{l},\textbf{k},-\textbf{k}}\right)
\texttt{Sp}\left\{\widehat{\varrho}_{0}C_{-\textbf{l},\gamma}^{+}(\tau_{1})C_{\textbf{l},\alpha}^{+}(\tau_{1})C_{\textbf{k},\alpha}(\tau_{1})C_{-\textbf{k},\gamma}(\tau_{1})\right\}
\nonumber\\
&&+\frac{1}{2!}\frac{1}{V^{2}}\int_{0}^{\beta}d\tau_{2}\int_{0}^{\beta}d\tau_{1}\sum_{\alpha,\gamma}\sum_{\textbf{k},\textbf{l}}\left(-\frac{1}{2}V_{\textbf{l},-\textbf{l},\textbf{k},-\textbf{k}}\right)
\sum_{\alpha',\gamma'}\sum_{\textbf{k}',\textbf{l}'}\left(-\frac{1}{2}V_{\textbf{l}',-\textbf{l}',\textbf{k}',-\textbf{k}'}\right)\\
&&\times\texttt{Sp}\left\{\widehat{\varrho}_{0}C_{-\textbf{l}',\gamma'}^{+}(\tau_{2})
C_{\textbf{l}',\alpha'}^{+}(\tau_{2})C_{\textbf{k}',\alpha'}(\tau_{2})C_{-\textbf{k}',\gamma'}(\tau_{2})
C_{-\textbf{l},\gamma}^{+}(\tau_{1})C_{\textbf{l},\alpha}^{+}(\tau_{1})C_{\textbf{k},\alpha}(\tau_{1})C_{-\textbf{k},\gamma}(\tau_{1})\right\}+...,\nonumber
\end{eqnarray}
where we took into account that
$\texttt{Sp}\left\{\widehat{\varrho}_{0}\right\}=1$. In order to
calculate (\ref{1.21}) approximately we can uncouple correlations
by the following way taking into account anticommutation of
operators $C$ and $C^{+}$:
\begin{eqnarray}\label{1.22}
&&R(\beta)\approx
1+\frac{1}{1!}(-1)^{2}\frac{1}{V}\int_{0}^{\beta}d\tau_{1}\sum_{\alpha,\gamma}\sum_{\textbf{k},\textbf{l}}\left(-\frac{1}{2}V_{\textbf{l},-\textbf{l},\textbf{k},-\textbf{k}}\right)
\texttt{Sp}\left\{\widehat{\varrho}_{0}C_{\textbf{l},\alpha}^{+}(\tau_{1})C_{-\textbf{l},\gamma}^{+}(\tau_{1})\right\}
\texttt{Sp}\left\{\widehat{\varrho}_{0}C_{-\textbf{k},\gamma}(\tau_{1})C_{\textbf{k},\alpha}(\tau_{1})\right\}
\nonumber\\
&&+\frac{1}{2!}(-1)^{4}\frac{1}{V^{2}}\int_{0}^{\beta}d\tau_{2}\int_{0}^{\beta}d\tau_{1}\sum_{\alpha,\gamma}\sum_{\textbf{k},\textbf{l}}\left(-\frac{1}{2}V_{\textbf{l},-\textbf{l},\textbf{k},-\textbf{k}}\right)
\sum_{\alpha',\gamma'}\sum_{\textbf{k}',\textbf{l}'}\left(-\frac{1}{2}V_{\textbf{l}',-\textbf{l}',\textbf{k}',-\textbf{k}'}\right)\nonumber\\
&&\times\texttt{Sp}\left\{\widehat{\varrho}_{0}C_{\textbf{l}',\alpha'}^{+}(\tau_{2})C_{-\textbf{l}',\gamma'}^{+}(\tau_{2})\right\}\texttt{Sp}\left\{\widehat{\varrho}_{0}
C_{-\textbf{k}',\gamma'}(\tau_{2})C_{\textbf{k}',\alpha'}(\tau_{2})\right\}\texttt{Sp}\left\{\widehat{\varrho}_{0}
C_{\textbf{l},\alpha}^{+}(\tau_{1})C_{-\textbf{l},\gamma}^{+}(\tau_{1})\right\}\texttt{Sp}\left\{\widehat{\varrho}_{0}
C_{-\textbf{k},\gamma}(\tau_{1})C_{\textbf{k},\alpha}(\tau_{1})\right\}
\nonumber\\
&&+\ldots=1+R_{1}+\frac{1}{2!}R_{1}^{2}+\ldots=\exp(R_{1})
\end{eqnarray}
Let us take into account that our approximation is analogous to
Fock approximation for normal processes. A decay of
quasi-particles is absent in Hartree-Fock approximation, hence the
amplitude of pairing is real $\Delta=\Delta^{+}$ in the momentum
space. As  a consequence we have $F=-F^{+}$. Then $R(t)$ can be
written as
\begin{eqnarray}\label{1.23}
    \ln R(\beta)=R_{1}(\beta)&=&\frac{1}{V}\int_{0}^{\beta}d\tau_{1}\sum_{\alpha,\gamma}\sum_{\textbf{k},\textbf{l}}\left(-\frac{1}{2}V_{\textbf{l},-\textbf{l},\textbf{k},-\textbf{k}}\right)
iF_{\alpha\gamma}(\textbf{l},\tau_{1}-\tau_{1})
iF_{\alpha\beta}(\textbf{k},\tau_{1}-\tau_{1})\nonumber\\
&=&\frac{2}{V}\sum_{\textbf{k},\textbf{l}}\left(\frac{1}{2}V_{\textbf{l},-\textbf{l},\textbf{k},-\textbf{k}}\right)
F(\textbf{l},\tau\rightarrow 0^{-})F(\textbf{k},\tau\rightarrow 0^{-})\beta\nonumber\\
&=&\frac{\beta\lambda}{V}\sum_{\textbf{k}}w_{k}\frac{1}{\beta}\sum_{n=-\infty}^{n=+\infty}F(\textbf{k},\omega_{n})
\sum_{\textbf{k}}w_{k}\frac{1}{\beta}\sum_{n=-\infty}^{n=+\infty}F(\textbf{k},\omega_{n})\nonumber\\
&=&-\beta\lambda
V\nu_{F}^{2}\int_{-\omega_{D}}^{\omega_{D}}\tanh\left(\frac{\beta
E}{2}\right)\frac{\Delta}{2E}d\varepsilon\int_{-\omega_{D}}^{\omega_{D}}\tanh\left(\frac{\beta
E}{2}\right)\frac{\Delta}{2E}d\varepsilon.
\end{eqnarray}
\begin{figure}[h]
\includegraphics[width=8cm]{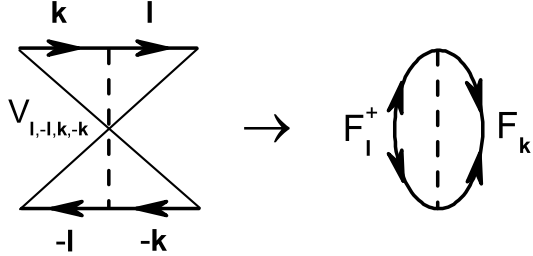}
\caption{The uncoupling of correlations in the vacuum amplitude
for the process of scattering of two fermions by the interaction
potential $V_{\textbf{l},-\textbf{l},\textbf{k},-\textbf{k}}$ from
the initial states $\textbf{k},\alpha$ and $-\textbf{k},\gamma$ to
the final states $\textbf{l},\alpha$ and $-\textbf{l},\gamma$. As
a result we have the anomalous transition amplitude
"vacuum-vacuum".} \label{fig2}
\end{figure}
The procedure of uncoupling of correlations and the contribution
of the anomalous process in the vacuum amplitude can be
represented graphically (Fig.\ref{fig2}). In order to calculate
the contribution of interaction into free energy we can use the
formula (\ref{1.12}):
\begin{eqnarray}\label{1.24}
\Omega_{\lambda}&=&-\frac{1}{\beta}\ln R(\beta)
=-\frac{\lambda}{V}\sum_{\textbf{k}}w_{k}\frac{1}{\beta}\sum_{n=-\infty}^{n=+\infty}F(\textbf{k},\omega_{n})
\sum_{\textbf{k}}w_{k}\frac{1}{\beta}\sum_{n=-\infty}^{n=+\infty}F(\textbf{k},\omega_{n})
\nonumber\\
&=&\lambda
V\nu_{F}^{2}\int_{-\omega_{D}}^{\omega_{D}}\tanh\left(\frac{\beta
E}{2}\right)\frac{\Delta}{2E}d\varepsilon\int_{-\omega_{D}}^{\omega_{D}}\tanh\left(\frac{\beta
E}{2}\right)\frac{\Delta}{2E}d\varepsilon.
\end{eqnarray}
If we suppose that $\Delta=0$, then we shall have
$\Omega_{\lambda}=0$.

\subsection{Free energy.}

Starting from the above found results we can write the expression
for free energy of a system:
\begin{eqnarray}\label{1.25}
&&\Omega=\langle
W\rangle-\frac{1}{\beta}S+\Omega_{\lambda}=-\frac{2i}{\beta}\lim_{\tau\rightarrow0^{-}}\sum_{\textbf{k}}\sum_{n=-\infty}^{n=+\infty}G(\textbf{k},\omega_{n})e^{-i\omega_{n}
t}\varepsilon(k)\nonumber\\
&&+\frac{2}{\beta}\sum_{\textbf{k}}\left[f(k)\ln f(k)+(1-f(k))\ln
(1-f(k))\right]
-\frac{\lambda}{V}\sum_{\textbf{k}}w_{k}\frac{1}{\beta}\sum_{n=-\infty}^{n=+\infty}F(\textbf{k},\omega_{n})
\sum_{\textbf{k}}w_{k}\frac{1}{\beta}\sum_{n=-\infty}^{n=+\infty}F(\textbf{k},\omega_{n})
\end{eqnarray}
The observed value of $\Delta$ must minimize the free energy:
\begin{eqnarray}\label{1.26}
\frac{d\Omega}{d\Delta}=0\Longrightarrow(-i)\Delta=\frac{\lambda}{V\beta}\sum_{\textbf{k}}\sum_{n=-\infty}^{n=+\infty}w_{k}F(\textbf{k},\omega_{n}).
\end{eqnarray}

The functional (\ref{1.25}) can be written in quadratures:
\begin{eqnarray}\label{1.27}
\Omega=&&\Omega_{n}+V\nu_{F}\int_{-\omega_{D}}^{\omega_{D}}\left[\tanh\left(\frac{\beta|\varepsilon|}{2}\right)\frac{\varepsilon^{2}}{|\varepsilon|}-
\tanh\left(\frac{\beta
E}{2}\right)\frac{\varepsilon^{2}}{E}\right]d\varepsilon\nonumber\\
&&+\frac{2V}{\beta}\nu_{F}\int_{-\omega_{D}}^{\omega_{D}}\left[f_{S}\ln
f_{S}+(1-f_{S})\ln (1-f_{S})-f_{0}\ln f_{0}-(1-f_{0})\ln
(1-f_{0})\right]d\varepsilon\nonumber\\
&&+V\nu_{F}g\int_{-\omega_{D}}^{\omega_{D}}\tanh\left(\frac{\beta
E}{2}\right)\frac{\Delta}{2E}d\varepsilon\int_{-\omega_{D}}^{\omega_{D}}\tanh\left(\frac{\beta
E}{2}\right)\frac{\Delta}{2E}d\varepsilon,
\end{eqnarray}
where $\Omega_{n}$ is the energy of a normal phase,
$g\equiv\lambda\nu_{F}$ is the effective interaction constant, $V$
is the volume of a system. We have $\Omega=\Omega_{n}$ for the
case $\Delta=0$. The equilibrium value of $\Delta$ is determined
by balance of kinetic energy, the entropy term and the energy of
interaction. The parameter $g$ can be expressed via critical
temperature $\beta_{C}$ with help of $\Delta(\beta_{C})=0$ as
following:
\begin{equation}\label{1.28}
    1=-g\int_{-\omega_{D}}^{\omega_{D}}\tanh\left(\frac{\beta_{C}|\varepsilon|}{2}\right)\frac{1}{2|\varepsilon|}d\varepsilon.
\end{equation}

Let us consider a low-temperature limit of the free energy
(\ref{1.27}): $\Delta\beta\gg 1$ at $T\rightarrow 0$. This means,
that the value $\Delta-\Delta_{0}$ can be chosen as the expansion
parameter, where $\Delta_{0}=\Delta(T=0)$ is the equilibrium value
of gap (amplitude of pairing) at zero temperature. Then the
low-temperature expansion has a form:
\begin{equation}\label{1.29}
    \Omega=\Omega_{n}+V\left(\alpha_{0}(T)+b_{0}(T)\Delta+d_{0}\Delta^{2}\right),
\end{equation}
where the expansion coefficients in the approximation $g\ll
1\Rightarrow\Delta\ll\omega_{D}$ are
\begin{eqnarray}\label{1.30}
\alpha_{0}(T) &=&
\frac{\nu_{F}}{2}\Delta_{0}^{2}+4\nu_{F}T^{2}-\nu_{F}\sqrt{8\pi\Delta_{0}T^{3}}e^{-\frac{\Delta_{0}}{T}}
-\nu_{F}\sqrt{8\pi\Delta_{0}^{3}T}e^{-\frac{\Delta_{0}}{T}} \nonumber\\
b_{0}(T) &=&
-2\nu_{F}\Delta_{0}+\nu_{F}\sqrt{8\pi\Delta_{0}T}e^{-\frac{\Delta_{0}}{T}},
\qquad  d_{0} = \nu_{F}
\end{eqnarray}
The equilibrium value of the energy gap is
$\Delta(T)=\frac{b_{0}(T)}{2d_{0}}=\Delta_{0}\left(1-\sqrt{\frac{2\pi
T}{\Delta_{0}}}e^{-\frac{\Delta_{0}}{T}}\right)$.

Let us consider a high-temperature limit of the free energy:
$\Delta\beta_{C}\ll 1$ at $T\rightarrow T_{C}$. Due to a rapid
convergence of integration elements in (\ref{1.27}), the limits of
integration can be $-\infty,+\infty$. Then the expansion in powers
of $\Delta$ gives:
\begin{equation}\label{1.32}
    \Omega=\Omega_{n}+V\left(\alpha(T)\Delta^{2}+\frac{1}{2}b\Delta^{4}+\frac{1}{3}d\Delta^{6}\right),
\end{equation}
where the coefficients of the expansion are
\begin{eqnarray}\label{1.33}
  \alpha(T) &=& \nu_{F}\frac{T-T_{c}}{T_{c}},\qquad
  b = \nu_{F}\frac{7\zeta(3)}{8\pi^{2}T_{c}^{2}}\\
  d &=&
\nu_{F}\left(\frac{52.31\zeta(5)}{\pi^{4}}+4.83\right)\frac{1}{4!T_{c}^{4}}\nonumber.
\end{eqnarray}
This expansion has a form of Landau expansion of free energy in
powers of the order parameter.

\section{Free energy of spatially inhomogeneous superconductor.}\label{spaceinhom}

In the previous sections we supposed, that amplitudes of pairing
$\Delta$ and $\Delta^{+}$ do not depend on spatial coordinates.
This takes place in interminable, homogeneous, isotropic and
isolated from external fields superconductor. However in a general
case these conditions are not realized. For example, in a
sufficiently strong magnetic field the inclusions of normal phase
can exist in volume of a superconductor. Another example is the
contact of a superconductor and a normal metal. In this case the
order parameter is suppressed in the boundary layer of a
superconductor, however it appears in the boundary layer of a
normal metal. We understand a spatially inhomogeneous
superconductor as a superconductor with spatially inhomogeneous
distribution of the order parameter: $\Delta(\textbf{r})$ and
$\Delta^{+}(\textbf{r})$.

In the previous sections we considered the pairing of fermions
with the opposite momentums: $\textbf{k}$ and $-\textbf{k}$. Now
let fermions can pair with arbitrary momentums:
$\textbf{k}+\textbf{q}$ and $-\textbf{k}$, where $\textbf{q}$
passes through all the vector space. The potential of particles'
interaction in the given states is
$V_{\textbf{l}+\textbf{q},-\textbf{l},\textbf{k}+\textbf{q},-\textbf{k}}=\lambda
w_{k}w_{l}, \qquad \lambda<0$. For convenience let us pass to the
reference system of center of mass of a pair, where the momentums
of particles are equal by modulus and are directed oppositely:
\begin{eqnarray}\label{2.01a}
&&\textbf{p}_{01}=\frac{m_{2}\textbf{p}_{1}-m_{1}\textbf{p}_{2}}{m_{1}m_{2}}=\textbf{k}+\frac{\textbf{q}}{2}\nonumber\\
&&\textbf{p}_{01}=-\frac{m_{2}\textbf{p}_{1}-m_{1}\textbf{p}_{2}}{m_{1}m_{2}}=-\textbf{k}-\frac{\textbf{q}}{2}
\end{eqnarray}
because $m_{1}=m_{2}=m$. Hence the system hamiltonian has the
form:
\begin{eqnarray}\label{2.01}
\widehat{H}_{\texttt{eff}}&=&\sum_{\textbf{q}}\sum_{\alpha}\sum_{\textbf{k}}\varepsilon\left(\textbf{k}+\frac{\textbf{q}}{2}\right)C_{\textbf{k}+\frac{\textbf{q}}{2},\alpha}^{+}C_{\textbf{k}+\frac{\textbf{q}}{2},\alpha}\nonumber\\
&+&\frac{1}{2V}\sum_{\textbf{q}}\sum_{\alpha,\gamma}\sum_{\textbf{k},\textbf{l}}\lambda
w_{k}w_{l}g_{\alpha\gamma}C_{-\textbf{l}-\frac{\textbf{q}}{2},\gamma}^{+}
C_{\textbf{l}+\frac{\textbf{q}}{2},\alpha}^{+}C_{\textbf{k}+\frac{\textbf{q}}{2},\alpha}C_{-\textbf{k}-\frac{\textbf{q}}{2},\gamma}.
\end{eqnarray}
In the system with a such hamiltonian the Cooper instability takes
place either as for the system with the hamiltonian (\ref{1.4}).
However now the amplitude of pairing is function of $\textbf{q}$:
$\Delta\rightarrow\Delta_{\textbf{q}}$. If the energy gap is
function of a wave vector $\Delta=\Delta(\textbf{q})$ then the
energy gap is function of a radius-vector:
$\Delta=\Delta(\textbf{r})$, because the Fourier transformation
realizes the one to one correspondence between functions defined
in the $\textbf{q}$-space and in the $\textbf{r}$-space. Taking
into account the reality of the order parameter in the momentum
space $\Delta_{\textbf{q}}=\Delta^{+}_{\textbf{q}}$ and requiring
identical dimension of the order parameter in the
$\textbf{q}$-space and in the $\textbf{r}$-space we can write the
Fourier-transformations as:
\begin{eqnarray}\label{2.1}
  \Delta(\textbf{r}) &=& \sum_{\textbf{q}}\Delta(\textbf{q})e^{i\textbf{qr}}=\frac{V}{(2\pi)^{3}}\int\Delta(\textbf{q})e^{i\textbf{qr}}d^{3}q, \nonumber\\
  \Delta^{+}(\textbf{r}) &=& \sum_{\textbf{q}}\Delta(q)e^{-i\textbf{qr}}=\frac{V}{(2\pi)^{3}}\int\Delta(\textbf{q})e^{-i\textbf{qr}}d^{3}q,\\
  \Delta(\textbf{q}) &=& \frac{1}{V}\int\Delta(\textbf{r})e^{-i\textbf{qr}}d^{3}r=\frac{1}{V}\int\Delta^{+}(\textbf{r})e^{i\textbf{qr}}d^{3}r.\nonumber
\end{eqnarray}

\begin{figure}[h]
\includegraphics[width=10cm]{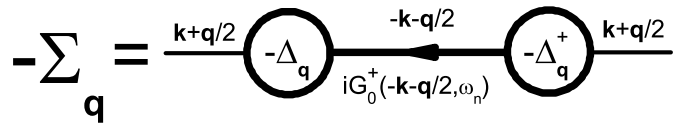}
\caption{The diagram for the mass operator $\Sigma$ describing an
interaction of a fermion with fluctuations of pairing in the
spatially inhomogeneous system.} \label{fig3}
\end{figure}

The mass operator for the process of interaction of a particle
with the fluctuations of pairing is shown in Fig.\ref{fig3}. In
analytical representation it has a view:
\begin{equation}\label{2.2}
    -\Sigma_{\textbf{q}}(\textbf{k},\omega_{n})=(-\Delta_{\textbf{q}})iG_{0}^{+}\left(-\textbf{k}-\frac{\textbf{q}}{2},\omega_{n}\right)(-\Delta^{+}_{\textbf{q}})
    =\frac{-\Delta_{\textbf{q}}\Delta^{+}_{\textbf{q}}}{i\omega_{n}+\varepsilon\left(\textbf{k}+\frac{\textbf{q}}{2}\right)},
\end{equation}
where the free propagator $G_{0}$ is
\begin{equation}\label{2.3}
    G_{0}=\frac{1}{i\omega_{n}-\varepsilon\left(\textbf{k}+\frac{\textbf{q}}{2}\right)}=
i\frac{i\omega_{n}+\varepsilon_{\textbf{q}}(k)}{(i\omega_{n})^{2}-\varepsilon^{2}_{\textbf{q}}(k)}.
\end{equation}
Then from Dyson equation we can obtain the dressed propagator:
\begin{eqnarray}\label{2.4}
\frac{1}{G_{0}}=\frac{1}{G_{S}}-i\Sigma_{\textbf{q}}\Rightarrow
G_{S}=i\frac{i\omega_{n}+\varepsilon_{\textbf{q}}}{(i\omega_{n})^{2}-E_{\textbf{q}}^{2}},
\end{eqnarray}
where $E_{\textbf{q}}$ is the specter of quasi-particles in a
inhomogeneous system:
\begin{equation}\label{2.5}
  E_{\textbf{q}}^{2}=\varepsilon_{\textbf{q}}^{2}+|\Delta_{\textbf{q}}|^{2},\qquad
\varepsilon_{\textbf{q}}\equiv\varepsilon\left(\textbf{k}+\frac{\textbf{q}}{2}\right)\approx\varepsilon(k)+\frac{\textbf{kq}}{2m},\qquad
|\textbf{k}|\simeq k_{F},\qquad |\textbf{q}|\ll k_{F}.
\end{equation}
Dyson equation can be represented in a form of Gor'kov equations
set, from where the expressions for anomalous propagators follow:
\begin{eqnarray}\label{2.6}
\begin{array}{c}
  (i\omega_{n}-\varepsilon_{\textbf{q}})G+\Delta_{\textbf{q}} F^{+}=i \\
\\
  (i\omega_{n}+\varepsilon_{\textbf{q}})F^{+}+G\Delta_{\textbf{q}}=0\\
\end{array}\Rightarrow
\begin{array}{c}
  F^{+}_{\textbf{q}}\left(\textbf{k},\omega_{n}\right)=\frac{-i\Delta_{\textbf{q}}^{+}}{(i\omega_{n})^{2}-E^{2}_{\textbf{q}}} \\
\\
  F_{\textbf{q}}\left(\textbf{k},\omega_{n}\right)=\left(F^{+}_{\textbf{q}}\left(\textbf{k},\omega_{n}\right)\right)^{+}=\frac{i\Delta_{\textbf{q}}}{(i\omega_{n})^{2}-E^{2}_{\textbf{q}}}\\
\end{array}
\end{eqnarray}
If to suppose $q=0$, then we shall have the expressions
(\ref{1.11}-\ref{1.11a}).

Now let us suppose that the order parameter $\Delta(\textbf{r})$
changes very slowly on a coherence length $l(T)$ which
characterizes a size of Cooper pair (the long-wave approximation
$q\rightarrow 0$, $ql(T)\ll 1$). Then we can suppose
$\varepsilon\left(\textbf{k}+\frac{\textbf{q}}{2}\right)\approx\varepsilon(\textbf{k})$
in the specter of quasi-particles, such that
$E_{\textbf{q}}\approx\sqrt{\varepsilon^{2}(k)+|\Delta_{\textbf{q}}|^{2}}$.
However we must keep
$\varepsilon\left(\textbf{k}+\frac{\textbf{q}}{2}\right)$ in
numerators of the expressions (\ref{2.3}) and (\ref{2.4}) for $G$.
Hence the normal propagator has a form:
\begin{eqnarray}\label{2.7}
  G_{\textbf{q}}(\textbf{k},\tau)=-i\theta_{\tau}\left(g^{+}_{\textbf{q}}A\left(\textbf{k}+\frac{\textbf{q}}{2}\right)e^{-E_{\textbf{q}}\tau}+g^{-}_{\textbf{q}}B\left(\textbf{k}+\frac{\textbf{q}}{2}\right)e^{E_{\textbf{q}}\tau}\right)+
i\theta_{-\tau}\left(g^{-}_{\textbf{q}}A\left(\textbf{k}+\frac{\textbf{q}}{2}\right)e^{-E_{\textbf{q}}\tau}+g^{+}_{\textbf{q}}B\left(\textbf{k}+\frac{\textbf{q}}{2}\right)e^{E_{\textbf{q}}\tau}\right),
\end{eqnarray}
where
\begin{eqnarray}\label{2.8}
A\left(\textbf{k}+\frac{\textbf{q}}{2}\right)\approx
A_{\textbf{q}}(k)+\frac{1}{2E_{\textbf{q}}}\frac{\textbf{kq}}{2m},
\qquad B\left(\textbf{k}+\frac{\textbf{q}}{2}\right)\approx
B_{\textbf{q}}(k)-\frac{1}{2E_{\textbf{q}}}\frac{\textbf{kq}}{2m}.
\end{eqnarray}
The anomalous propagators are
\begin{eqnarray}\label{2.9}
  F^{+}_{\alpha\gamma}(\textbf{k},\textbf{q},\tau)&=&ig_{\alpha\beta}\frac{\Delta^{+}_{\textbf{q}}}{2E_{\textbf{q}}}
\left[\left(g_{\textbf{q}}^{+}e^{-E_{\textbf{q}}\tau}-g_{\textbf{q}}^{-}e^{E_{\textbf{q}}\tau}\right)\theta_{\tau}-\left(g_{\textbf{q}}^{+}e^{E_{\textbf{q}}\tau}-g_{\textbf{q}}^{-}e^{-E_{\textbf{q}}\tau}\right)\theta_{-\tau}\right],\nonumber\\
F_{\alpha\gamma}(\textbf{k},\textbf{q},\tau)&=&ig_{\alpha\beta}\frac{\Delta_{\textbf{q}}}{2E_{\textbf{q}}}
\left[-\left(g_{\textbf{q}}^{+}e^{-E_{\textbf{q}}\tau}-g_{\textbf{q}}^{-}e^{E_{\textbf{q}}\tau}\right)\theta_{\tau}+\left(g_{\textbf{q}}^{+}e^{E_{\textbf{q}}\tau}-g_{\textbf{q}}^{-}e^{-E_{\textbf{q}}\tau}\right)\theta_{-\tau}\right].
\end{eqnarray}
We can see, that in the long wave approximation the anomalous
propagators depend on
 $\textbf{q}$ by means of $\Delta(\textbf{q})$ only.

Kinetic energy of a system is determined by the following way:
\begin{eqnarray}\label{2.10}
\langle
W\rangle&=&-2i\sum_{\textbf{q}}\sum_{\textbf{k}}\varepsilon\left(\textbf{k}+\frac{\textbf{q}}{2}\right)G\left(\textbf{k}+\frac{\textbf{q}}{2},\tau\rightarrow 0^{-}\right)\nonumber\\
&=&W_{n}+V\nu_{F}\sum_{\textbf{q}}\int_{-\omega_{D}}^{\omega_{D}}\left[\tanh\left(\frac{\beta|\varepsilon|}{2}\right)\frac{\varepsilon^{2}}{|\varepsilon|}-
\tanh\left(\frac{\beta
E_{\textbf{q}}}{2}\right)\frac{\varepsilon^{2}}{E_{\textbf{q}}}\right]d\varepsilon\nonumber\\
&+&V\nu_{F}\frac{v_{F}^{2}}{12}\sum_{\textbf{q}}\int_{-\omega_{D}}^{\omega_{D}}\textbf{q}^{2}\left[\tanh\left(\frac{\beta|\varepsilon|}{2}\right)\frac{1}{|\varepsilon|}-
\tanh\left(\frac{\beta
E_{\textbf{q}}}{2}\right)\frac{1}{E_{\textbf{q}}}\right]d\varepsilon.
\end{eqnarray}
We can see, that the term, which is proportional to
$\textbf{q}^{2}$, is added to the kinetic energy (\ref{1.16})
(with the replacement $E\rightarrow
E_{\textbf{q}}=\sqrt{\varepsilon^{2}(k)+|\Delta(\textbf{q})|^{2}}$).
In the long wave approximation the expressions for entropy and
vacuum amplitude coincide with the expressions (\ref{1.18}) and
(\ref{1.23}) accordingly, however it should be written
$\Delta(\textbf{q})$ instead of $\Delta=\texttt{const}$. Then we
can write the free energy:
\begin{equation}\label{2.11}
    \Omega=\Omega_{n}+\sum_{\textbf{q}}\Omega(\Delta_{\textbf{q}})+V\nu_{F}\frac{v_{F}^{2}}{12}\sum_{\textbf{q}}\int_{-\omega_{D}}^{\omega_{D}}\textbf{q}^{2}\left[\tanh\left(\frac{\beta|\varepsilon|}{2}\right)\frac{1}{|\varepsilon|}-
\tanh\left(\frac{\beta
E_{\textbf{q}}}{2}\right)\frac{1}{E_{\textbf{q}}}\right]d\varepsilon,
\end{equation}
where $\Omega(\Delta_{\textbf{q}})$ coincides with the expression
(\ref{1.27}), where the replacement
$\Delta\rightarrow\Delta(\textbf{q})$ was done.

Expanding the free energy (\ref{2.11}) in powers of $\Delta$ we
can obtain the expression:
\begin{equation}\label{2.12}
    \Omega=\Omega_{n}+V\sum_{\textbf{q}}\left(\alpha(T)\Delta_{\textbf{q}}^{2}+\frac{1}{2}b\Delta_{\textbf{q}}^{4}+\gamma q^{2}\Delta_{\textbf{q}}^{2}\right)
    =\Omega_{n}+\frac{V^{2}}{(2\pi)^{3}}\int\left(\alpha(T)\Delta_{\textbf{q}}^{2}+\frac{1}{2}b\Delta_{\textbf{q}}^{4}+\gamma
\textbf{q}^{2}\Delta_{\textbf{q}}^{2}\right)d^{3}q,
\end{equation}
where the coefficient $\gamma$ is
\begin{eqnarray}\label{2.13}
 \gamma=\nu_{F}\frac{7\zeta(3)v_{F}^{2}}{48\pi^{2}T_{c}^{2}}=\nu_{F}l_{0}^{2},
\end{eqnarray}
where $l_{0}$ is a coherence length at $T=0$ (Pippard length). The
expansion (\ref{2.12}) has a form of Landau expansion of free
energy in powers of order parameter at the condition $ql_{0}\ll
1$. We can see, that a spatial inhomogeneity increases the free
energy of a superconductor. Hence in most cases we can restrict
the free energy expansion to the term $\sim q^{2}$, because more
fast changes of $\Delta$ increase the free energy essentially.

Let's pass from momentum space to real space using the expressions
(\ref{2.1}):
\begin{eqnarray}\label{2.15}
\int\Delta_{\textbf{q}}\Delta_{\textbf{q}}d^{3}q=\int\Delta_{\textbf{q}}\left[\frac{1}{V}\int\Delta(\textbf{r})e^{-i\textbf{qr}}d^{3}r\right]d^{3}q
=\frac{1}{V}\int\Delta(\textbf{r})\left[\int\Delta(\textbf{q})e^{-i\textbf{qr}}d^{3}q\right]d^{3}r
=\frac{(2\pi)^{3}}{V^{2}}\int\Delta(\textbf{r})\Delta^{+}(\textbf{r})d^{3}r,
\end{eqnarray}
\begin{eqnarray}\label{2.16}
&&\int\textbf{q}\Delta_{\textbf{q}}\textbf{q}\Delta_{\textbf{q}}d^{3}q=\int\textbf{q}\Delta_{\textbf{q}}\left[\frac{1}{V}\int
e^{-i\textbf{qr}}(-i)\frac{\partial}{\partial\textbf{r}}\Delta(\textbf{r})d^{3}r\right]d^{3}q\nonumber\\
&&=\frac{-i}{V}\int\left[\int\textbf{q}\Delta(\textbf{q})e^{-i\textbf{qr}}d^{3}q\right]\frac{\partial}{\partial\textbf{r}}\Delta(\textbf{r})d^{3}r
=\frac{(2\pi)^{3}}{V^{2}}\int\left[\frac{\partial}{\partial\textbf{r}}\Delta(\textbf{r})\right]\frac{\partial}{\partial\textbf{r}}\Delta^{+}(\textbf{r})d^{3}r.
\end{eqnarray}
For the term $\Delta^{4}$ and terms with more high powers the
situation is more difficult. This is because a square of a Fourier
transform is not equal to a Fourier transform of a square:
$\left(\frac{1}{V}\int\Delta(\textbf{r})e^{-i\textbf{qr}}d^{3}r\right)^{2}\neq\frac{1}{V}\int\Delta^{2}(\textbf{r})e^{-i\textbf{qr}}d^{3}r$.
Apparently this fact is the manifestation of a nonlocality of the
order parameter in zero magnetic field \citep{hook}: the value of
gap in a point is determined by a distribution of the gap in a
some neighborhood of this point: $\Delta(\textbf{r})=\int
d\textbf{r}'Q(\textbf{r},\textbf{r}')\Delta(\textbf{r}')$, where
$Q\left(\textbf{r},\textbf{r}'\right)$ is the kernel in coordinate
space. In a simplest case the kernel $Q$ is the function of the
distance $|\textbf{r}-\textbf{r}'|$ only. In a total nonlocal case
the kernel is function of the gap $\Delta(\textbf{r}')$ also. If
we neglect of this correlation then we shall obtain the expansion
of the free energy in powers of $\Delta\Delta^{+}$ in real space
in the following form:
\begin{eqnarray}\label{2.17}
\Omega=\Omega_{n}+\int\left[\alpha(T)|\Delta(\textbf{r})|^{2}+\frac{b}{2}|\Delta(\textbf{r})|^{4}
+\gamma\left|\frac{\partial}{\partial\textbf{r}}\Delta(\textbf{r})\right|^{2}\right]d^{3}r.
\end{eqnarray}
This expansion coincides with Ginzburg-Landau expansion in zero
magnetic field. As a result of minimization of the functional
(\ref{2.17}) $\frac{\delta\Omega}{\delta\Delta}=0$ with a
consideration of corresponding boundary conditions we obtain the
energy gap as the function of coordinates
$\Delta=\Delta(\textbf{r})$. It prove our statement about that the
pairing of fermions with the momentums $\textbf{k}+\textbf{q}$ and
$-\textbf{k}$ (or $\textbf{k}+\frac{\textbf{q}}{2}$ and
$-\textbf{k}-\frac{\textbf{q}}{2}$ in a system of center of mass)
results to the spatially inhomogeneous order parameter if the
wave-vector $\textbf{q}$ passes through all the vector space. The
physical interpretation of this fact lies in the following. Let us
assume that we have a superconductor with spatially homogeneous
distribution of the order parameter. This distribution corresponds
to a minimum of the free energy. Then we shall create an
inhomogeneity in the distribution of energy gap by some way. This
inhomogeneity increases the free energy according to (\ref{2.11})
or (\ref{2.17}). This means that we is doing some work on the
system. This work is spent on infusion of an additional momentum
to each partner of a pair:
$\textbf{k}\rightarrow\textbf{k}+\frac{\textbf{q}}{2}$ and
$-\textbf{k}\rightarrow-\textbf{k}-\frac{\textbf{q}}{2}$.

Minimization of the functional (\ref{2.12}) gives (at the
condition $q\ll l(T)$):
\begin{eqnarray}\label{2.15a}
\Delta_{\textbf{q}}^{2}=-\frac{\alpha}{b}\left(1+\frac{\gamma}{\alpha}q^{2}\right)=\Delta^{2}(T)\left(1-l^{2}(T)q^{2}\right)
\approx\frac{\Delta^{2}(T)}{1+l^{2}(T)q^{2}} \Rightarrow
Q(\textbf{q})\approx\frac{1}{1+l^{2}(T)q^{2}/2},
\end{eqnarray}
where $Q(\textbf{q})$ is the kernel in momentum space,
$l(T)=l_{0}/|\alpha(T)|$ is the coherent length at temperature
$T$. At the critical temperature $T\rightarrow T_{C}$ we have
$l(T_{C})=\infty$. This means that at high temperatures the
nonlocality is caused by both the own nonlocality of a
superconductor and by fluctuations in the critical region. We can
see that in the long-wave approximation the kernel $Q$ is the
function of the distance $|\textbf{r}-\textbf{r}'|$ only:
$Q\left(|\textbf{r}-\textbf{r}'|\right)=\sum_{\textbf{q}}Q(\textbf{q})e^{i\textbf{q}(\textbf{r}-\textbf{r}')}$.

Let's consider the low-temperature expansion:
\begin{equation}\label{2.18}
\Omega=\Omega_{n}+V\sum_{\textbf{q}}\left(\alpha_{0}(T)+b_{0}(T)\Delta_{\textbf{q}}+d_{0}\Delta^{2}_{\textbf{q}}+\gamma
q^{2}\Delta_{\textbf{q}}^{2}\right),
\end{equation}
where coefficients $\alpha_{0}(T),b_{0}(T),d_{0}$ are determined
by the formulas (\ref{1.30}), and coefficient $\gamma$ is
determined by the formula (\ref{1.29}). The observed configuration
of order parameter $\Delta_{\textbf{q}}$ minimized the free
energy:
\begin{eqnarray}\label{2.19}
  \frac{\delta\Omega}{\delta\Delta}=0 &\Rightarrow& b_{0}(T)+2d_{0}\Delta_{\textbf{q}}+\gamma q^{2}\Delta_{\textbf{q}}=0
  \Rightarrow \Delta_{\textbf{q}}(T)=\frac{-b_{0}(T)}{2d_{0}+2\gamma
  q^{2}}=\frac{\Delta(T)}{1+l_{0}^{2}q^{2}/2}\Rightarrow Q\approx\frac{1}{1+l_{0}^{2}q^{2}/2}
\end{eqnarray}
Thus the energy gap (the amplitude of pairing)
$\Delta_{\textbf{q}}$ is the nonlocal order parameter with the
correlation length $l_{0}$.

The dependence of the interaction constant on coordinates
$\lambda(\textbf{r})=
\sum_{\textbf{q}}\lambda(\textbf{q})e^{i\textbf{qr}}$ is the
direct cause of a spatial inhomogeneity. Hamiltonian of such
system is
\begin{eqnarray}\label{2.20b}
\widehat{H}_{\texttt{eff}}&=&\sum_{\textbf{q}}\sum_{\alpha}\sum_{\textbf{k}}\varepsilon_{q}
\left(\textbf{k}+\frac{\textbf{q}}{2}\right)C_{\textbf{k}+\frac{\textbf{q}}{2},\alpha}^{+}C_{\textbf{k}+\frac{\textbf{q}}{2},\alpha}\nonumber\\
&+&\frac{1}{2V}\sum_{\textbf{q}}\sum_{\alpha,\gamma}\sum_{\textbf{k},\textbf{l}}\lambda(\textbf{q})w_{k}w_{l}g_{\alpha\gamma}
C_{-\textbf{l}-\frac{\textbf{q}}{2},\gamma}^{+}
C_{\textbf{l}+\frac{\textbf{q}}{2},\alpha}^{+}C_{\textbf{k}+\frac{\textbf{q}}{2},\alpha}C_{-\textbf{k}-\frac{\textbf{q}}{2},\gamma},
\end{eqnarray}
Hence the free energy of a spatially inhomogeneous superconductor
has the form:
\begin{eqnarray}\label{2.20}
&&\Omega=-\frac{2i}{\beta}\lim_{\tau\rightarrow0^{-}}\sum_{\textbf{q}}\sum_{\textbf{k}}\sum_{n
=-\infty}^{n=+\infty}G_{\textbf{q}}\left(\textbf{k},\omega_{n}\right)e^{-i\omega_{n}
\tau}\varepsilon\left(\textbf{k}+\frac{\textbf{q}}{2}\right)\nonumber\\
&&+\sum_{\textbf{q}}\sum_{\textbf{k}}\frac{2}{\beta}\left[f_{\textbf{q}}^{S}\ln
f_{\textbf{q}}^{S}+(1-f_{\textbf{q}}^{S})\ln
(1-f_{\textbf{q}}^{S})-f_{\textbf{q}}^{0}\ln
f_{\textbf{q}}^{0}-(1-f_{\textbf{q}}^{0})\ln
(1-f_{\textbf{q},\textbf{a}}^{0})\right]\nonumber\\
&&-\sum_{\textbf{q}}\frac{\lambda(\textbf{q})}{V}\sum_{\textbf{k}}w_{k}\frac{1}{\beta}\sum_{n=-\infty}^{n=+\infty}F_{\textbf{q}}(\textbf{k},\omega_{n})
\sum_{\textbf{k}}w_{k}\frac{1}{\beta}\sum_{n=-\infty}^{n=+\infty}F_{\textbf{q}}(\textbf{k},\omega_{n})\
\end{eqnarray}
the observed value of $\Delta_{\textbf{q},\textbf{a}}$ must
minimize the free energy:
\begin{eqnarray}\label{2.20a}
\frac{d\Omega}{d\Delta_{\textbf{q}}}=0&&\Longrightarrow
(-i)\Delta_{\textbf{q}}=\frac{\lambda(\textbf{q})}{V\beta}\sum_{\textbf{k}}\sum_{n=-\infty}^{n=+\infty}w_{k}F_{\textbf{q}}(\textbf{k},\omega_{n})
\Rightarrow
1=-\frac{\lambda(\textbf{q})}{V}\sum_{\textbf{k}}w_{k}\frac{1}{2E_{\textbf{q}}}\tanh\left(\frac{\beta
E_{\textbf{q}}}{2}\right).
\end{eqnarray}
The equation (\ref{2.20a}) generalizes Ginzburg-Landau equation
for any temperature and for arbitrary spatial variations of the
order parameter. The kernel $Q$ is the function of both the
distance $|\textbf{r}-\textbf{r}'|$ and the order parameter
$\Delta(\textbf{r}')$ now. This means, that the fast changing in
space order parameter is strongly nonlocal.

In order to write the high-temperature expansion we can introduce
the local critical temperature $T_{C}(\textbf{q})$ with help of
the equation (\ref{2.20a}) assuming $\Delta=0$ and
$\varepsilon_{\textbf{q}}=\varepsilon$:
\begin{equation}\label{2.21}
    1=-g(\textbf{q})\int_{-\omega_{D}}^{\omega_{D}}\tanh\left(\frac{\beta_{C}(\textbf{q})|\varepsilon|}{2}\right)\frac{d\varepsilon}{2|\varepsilon|},
\end{equation}
where $g(\textbf{q})\equiv\lambda(\textbf{q})\nu_{F}$. If
$\Delta_{\textbf{q}}/T\ll 1$ then we can expanse the free energy
functional (\ref{2.20}) in powers of $\Delta_{\textbf{q}}$. In the
long-wave limit we has the expression:
\begin{equation}\label{2.22}
    \Omega=\Omega_{n}+V\sum_{\textbf{q}}\left(\nu_{F}\frac{T-T_{C}(\textbf{q})}{T_{C}(\textbf{q})}\Delta_{\textbf{q}}^{2}
    +\frac{\nu_{F}}{2}\frac{7\zeta(3)}{8\pi^{2}T_{C}^{2}(\textbf{q})}\Delta_{\textbf{q}}^{4}
    +\nu_{F}\frac{7\zeta(3)v_{F}^{2}}{48\pi^{2}T_{C}^{2}(\textbf{q})}
    q^{2}\Delta_{\textbf{q}}^{2}\right).
\end{equation}
Moreover the correlation length can be introduced formally:
$\gamma(\textbf{q})=\nu_{F}l_{0}^{2}(\textbf{q})$. In the
$\textbf{r}$-space the coefficients of the expansion is functions
of coordinates
$\alpha(T,\textbf{r}),b(\textbf{r}),\gamma(\textbf{r})$ and they
can be obtained by the microscopic way only.

The value $\Delta_{0}(\textbf{q})-\Delta_{\textbf{q}}(T)$ can be
the parameter of the low temperature expansion, where the value
$\Delta_{0}(\textbf{q})$ can be obtained from the equation
(\ref{2.20a}) assuming $T=0$ and
$\varepsilon_{\textbf{q}}=\varepsilon$:
\begin{eqnarray}\label{2.23}
1=-g(\textbf{q})\int_{-\omega_{D}}^{\omega_{D}}\frac{d\varepsilon}{2\sqrt{\varepsilon^{2}+\Delta_{0}^{2}(\textbf{q})}}\tanh\left(\frac{\beta
\sqrt{\varepsilon^{2}+\Delta_{0}^{2}(\textbf{q})}}{2}\right).
\end{eqnarray}
Then the low-temperature expansion has the form (\ref{2.18}) in
the long-wave limit, where the coefficients
$\alpha_{0}(T,\textbf{q}),b_{0}(\textbf{q})(T),d_{0}$ determined
by the formulas (\ref{1.30}) with the replacement
$\Delta_{0}\rightarrow\Delta_{0}(\textbf{q})$, and coefficient
$\gamma$ is determined by the formula (\ref{1.29}) with the
replacement $l_{0}\rightarrow l_{0}(\textbf{q})$. In the
$\textbf{r}$-space the coefficients is functions of coordinates
$\alpha_{0}(T,\textbf{r}),b_{0}(\textbf{r})(T),\gamma(\textbf{r})$
and they can be obtained by the microscopic way only.

\section{The free energy of spatially inhomogeneous superconductor in a magnetic field.}\label{magnetic}

In this section we shall generalize the previous results for the
case, when a superconductor is placed in magnetic field
$\textbf{H}(\textbf{r})=\texttt{rot}\textbf{A}(r)$. Our aim is to
obtain the functional of free energy
$\Omega\left(\Delta(\textbf{r}),\frac{\partial}{\partial\textbf{r}}\Delta(\textbf{r}),\textbf{A}(\textbf{r})\right)$,
which is correct for arbitrary value of the relation
$\Delta(T)/T$, for arbitrary spatial variations of the order
parameter $\Delta(\textbf{r})$, for an arbitrary value of a
magnetic penetration depth $\lambda(T)$ in comparison with a
coherent length $l_{0}$ (nonlocal electromagnetic response). Thus,
the all three restrictions on Ginzburg-Landau functional,
described in section \ref{intr}, are excluded.

Let the microscopic magnetic field with a potential
$\textbf{A}(\textbf{r})$ (with an intensity
$\textbf{H}(\textbf{r})$) exists in the given point of a
superconductor:
\begin{equation}\label{3.1}
\textbf{A}(\textbf{r})=\sum_{\textbf{q}}\textbf{a}_\textbf{q}e^{i\textbf{qr}}\Rightarrow
\textbf{H}(\textbf{r})=\texttt{rot}\textbf{A}(\textbf{r})=i\sum_{\textbf{q}}\textbf{q}\times\textbf{a}_\textbf{q}e^{i\textbf{qr}}.
\end{equation}
Moreover $\textbf{a}_{-\textbf{q}}=\textbf{a}_{\textbf{q}}^{+}$
follows from the reality of $\textbf{A}$. The energy of magnetic
field is
\begin{equation}\label{3.3}
    W_{f}=\frac{1}{8\pi}\int\left|\textbf{H}(\textbf{r})\right|^{2}d^{3}r
    =\frac{V}{8\pi}\sum_{\textbf{q}}\left(q^{2}(\textbf{a}_\textbf{q}\cdot\textbf{a}^{+}_\textbf{q})
    -(\textbf{q}\cdot\textbf{a}_\textbf{q})(\textbf{q}\cdot\textbf{a}^{+}_\textbf{q})\right).
\end{equation}
The magnetic field affects on a superconductor essentially. In the
first place, the distribution of order parameter becomes
inhomogeneous. As it was shown in the section \ref{spaceinhom},
the inhomogeneity leads to the growth of momentum of each element
of a pair: $\textbf{k}\rightarrow\textbf{k}+\frac{\textbf{q}}{2}$,
$-\textbf{k}\rightarrow -\textbf{k}-\frac{\textbf{q}}{2}$,
moreover the order parameter depends on the momentum
$\Delta=\Delta(\textbf{q})$. In the second place, the ordinary
momentum must be replaced by the canonical momentum:
$\textbf{k}+\frac{\textbf{q}}{2}\rightarrow\textbf{k}+\frac{\textbf{q}}{2}-\frac{e}{c}\textbf{a}_\textbf{q}$
and
$-\textbf{k}-\frac{\textbf{q}}{2}\rightarrow-\textbf{k}-\frac{\textbf{q}}{2}+\frac{e}{c}\textbf{a}_{-\textbf{q}}$,
moreover the order parameter is function of the momentums:
$\Delta=\Delta(\textbf{q},\textbf{a}_\textbf{q})$. For convenience
let's pass to the reference system of center of mass of a pair,
where the momentums of particles are equal by modulus, are
directed oppositely and are real:
\begin{eqnarray}\label{3.1a}
&&\textbf{p}_{01}=\frac{m_{2}\textbf{p}_{1}-m_{1}\textbf{p}_{2}}{m_{1}m_{2}}
=\textbf{k}+\frac{\textbf{q}}{2}-\frac{e}{2c}(\textbf{a}_\textbf{q}+\textbf{a}_\textbf{q}^{+})\nonumber\\
&&\textbf{p}_{01}=-\frac{m_{2}\textbf{p}_{1}-m_{1}\textbf{p}_{2}}{m_{1}m_{2}}
=-\textbf{k}-\frac{\textbf{q}}{2}+\frac{e}{2c}(\textbf{a}_\textbf{q}+\textbf{a}_\textbf{q}^{+})
\end{eqnarray}
because $m_{1}=m_{2}=m$. The hamiltonian of the system has a form:
\begin{eqnarray}\label{3.4a}
&&\widehat{H}_{\texttt{eff}}=\sum_{\textbf{q}}\sum_{\alpha}\sum_{\textbf{k}}
\varepsilon_{q}\left(\textbf{k}+\frac{\textbf{q}}{2}-\frac{e}{2c}(\textbf{a}_\textbf{q}+\textbf{a}_\textbf{q}^{+})\right)
C_{\textbf{k}+\frac{\textbf{q}}{2}-\frac{e}{2c}(\textbf{a}_\textbf{q}+\textbf{a}_\textbf{q}^{+}),\alpha}^{+}
C_{\textbf{k}+\frac{\textbf{q}}{2}-\frac{e}{2c}(\textbf{a}_\textbf{q}+\textbf{a}_\textbf{q}^{+}),\alpha}\nonumber\\
&&\\
&&+\frac{1}{2V}\sum_{\textbf{q}}\sum_{\alpha,\gamma}\sum_{\textbf{k},\textbf{l}}\lambda(\textbf{q})
w_{k}w_{l}g_{\alpha\gamma}C_{-\textbf{l}-\frac{\textbf{q}}{2}+\frac{e}{2c}(\textbf{a}_\textbf{q}+\textbf{a}_\textbf{q}^{+}),\gamma}^{+}
C_{\textbf{l}+\frac{\textbf{q}}{2}-\frac{e}{2c}(\textbf{a}_\textbf{q}+\textbf{a}_\textbf{q}^{+}),\alpha}^{+}
C_{\textbf{k}+\frac{\textbf{q}}{2}-\frac{e}{2c}(\textbf{a}_\textbf{q}+\textbf{a}_\textbf{q}^{+}),\alpha}
C_{-\textbf{k}-\frac{\textbf{q}}{2}+\frac{e}{2c}(\textbf{a}_\textbf{q}+\textbf{a}_\textbf{q}^{+}),\gamma},\nonumber
\end{eqnarray}
if $\lambda(\textbf{q})<0$. The hamiltonian (\ref{3.4a}) is the
generalization of BCS hamiltonian (\ref{1.4}) to the cases of the
dependence of the interaction constant on coordinates and to the
presence of a magnetic field.

\begin{figure}[h]
\includegraphics[width=18cm]{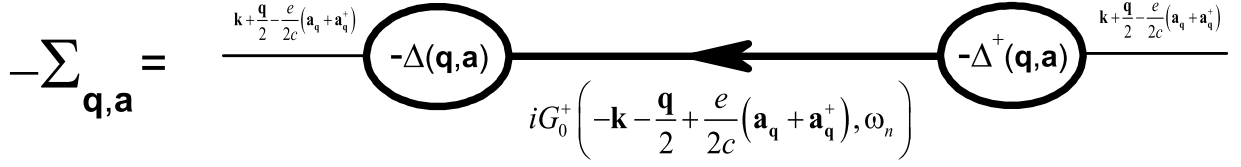}
\caption{The diagram for the mass operator $\Sigma$ describing the
interaction of a charged fermion with fluctuations of pairing in a
spatially inhomogeneous system situated in a magnetic field with
potential $\textbf{a}(\textbf{q})$.} \label{fig4}
\end{figure}

The system with the hamiltonian (\ref{3.4a}) is unstable regarding
in Cooper pairing. The mass operator for process of interaction of
a fermion with fluctuations of pairing is shown in Fig.\ref{fig4}.
In analytical representation this diagram has the form:
\begin{eqnarray}\label{3.4}
-\Sigma_{\textbf{q},\textbf{a}}\left(\textbf{k},\omega_{n}\right)=
-\Delta_{\textbf{q},\textbf{a}}iG_{0}^{+}\left(-\textbf{k}-\frac{\textbf{q}}{2}+\frac{e}{2c}(\textbf{a}_\textbf{q}+\textbf{a}_\textbf{q}^{+}),\omega_{n}\right)\left(-\Delta^{+}_{\textbf{q},\textbf{a}}\right)
=\frac{-\Delta_{\textbf{q},\textbf{a}}\Delta^{+}_{\textbf{q},\textbf{a}}}{i\omega_{n}+\varepsilon\left(-\textbf{k}-\frac{\textbf{q}}{2}+\frac{e}{2c}(\textbf{a}_\textbf{q}+\textbf{a}_\textbf{q}^{+})\right)},
\end{eqnarray}
where the free propagator $G_{0}$ is
\begin{equation}\label{3.5}
    G_{0}=\frac{1}{i\omega_{n}-\varepsilon\left(\textbf{k}+\frac{\textbf{q}}{2}-\frac{e}{2c}(\textbf{a}_\textbf{q}+\textbf{a}_\textbf{q}^{+})\right)}=
i\frac{i\omega_{n}+\varepsilon\left(\textbf{k}+\frac{\textbf{q}}{2}-\frac{e}{2c}(\textbf{a}_\textbf{q}+\textbf{a}_\textbf{q}^{+})\right)}
{(i\omega_{n})^{2}-\varepsilon^{2}\left(\textbf{k}+\frac{\textbf{q}}{2}-\frac{e}{2c}(\textbf{a}_\textbf{q}+\textbf{a}_\textbf{q}^{+})\right)}.
\end{equation}
From Dyson equation we can obtain the dressed propagator:
\begin{eqnarray}\label{3.6}
  \frac{1}{G_{0}} &=& \frac{1}{G_{S}}-i\Sigma_{\textbf{q}}\Rightarrow
G_{S}=i\frac{i\omega_{n}+\varepsilon\left(\textbf{k}+\frac{\textbf{q}}{2}-\frac{e}{2c}(\textbf{a}_\textbf{q}+\textbf{a}_\textbf{q}^{+})\right)}
{(i\omega_{n})^{2}-E^{2}_{\textbf{q},\textbf{a}}},
\end{eqnarray}
where $E$ is the specter of quasi-particles in an inhomogeneous
system situated in magnetic field:
\begin{eqnarray}\label{3.7}
&&E^{2}_{\textbf{q},\textbf{a}}
=\varepsilon^{2}\left(\textbf{k}+\frac{\textbf{q}}{2}-\frac{e}{2c}(\textbf{a}_\textbf{q}+\textbf{a}_\textbf{q}^{+})\right)
+|\Delta_{\textbf{q},\textbf{a}}|^{2}\nonumber\\
&&\varepsilon\left(\textbf{k}+\frac{\textbf{q}}{2}-\frac{e}{2c}(\textbf{a}_\textbf{q}+\textbf{a}_\textbf{q}^{+})\right)
\approx\varepsilon(k)+\frac{\textbf{k}\cdot\left(\frac{\textbf{q}}{2}-\frac{e}{2c}(\textbf{a}_\textbf{q}+\textbf{a}_\textbf{q}^{+})\right)}{m}\equiv
\varepsilon_{\textbf{q},\textbf{a}}, \qquad |\textbf{k}|\simeq
k_{F},
\end{eqnarray}
where we have introduced the notations $E_{\textbf{q},\textbf{a}}$
and $\varepsilon_{\textbf{q},\textbf{a}}$ for convenience. Dyson
equation can be represented in the form of Gor'kov equations. From
these equations the expressions for anomalous propagators follow:
\begin{eqnarray}\label{3.8}
\begin{array}{c}
  (i\omega_{n}-\varepsilon_{\textbf{q},\textbf{a}})G+\Delta_{\textbf{q},\textbf{a}} F^{+}=i \\
\\
  (i\omega_{n}+\varepsilon_{\textbf{q},\textbf{a}})F^{+}+G\Delta_{\textbf{q},\textbf{a}}=0\\
\end{array}\Rightarrow
\begin{array}{c}
  F^{+}_{\textbf{q},\textbf{a}}=\frac{-i\Delta_{\textbf{q},\textbf{a}}^{+}}{(i\omega_{n})^{2}-E^{2}_{\textbf{q},\textbf{a}}} \\
\\
  F_{\textbf{q},\textbf{a}}=(F^{+}_{\textbf{q},\textbf{a}})^{+}=\frac{i\Delta_{\textbf{q},\textbf{a}}}{(i\omega_{n})^{2}-E^{2}_{\textbf{q},\textbf{a}}}\\
\end{array}
\end{eqnarray}
If to suppose $\textbf{q}=0$ and $\textbf{a}=0$, then we shall
have the expressions (\ref{1.11}-\ref{1.12}).

Free energy of a superconductor is a sum of the following terms:
\begin{equation}\label{3.14}
    \Omega=\Omega_{n}+W_{S}-\frac{1}{\beta}S_{S}+\Omega_{\lambda}+W_{\texttt{f}}(\textbf{a}),
\end{equation}
where $W_{S}$ is the kinetic energy of fermions in superconducting
phase, $S_{S}$ is the entropy of boholons,
$\Omega_{\lambda}=-\frac{1}{\beta}\ln R(\beta)$ is the energy
corresponding to an interaction, $W_{f}(\textbf{a})$ is the energy
of magnetic field (\ref{3.3}). The expressions for $W_{S}$, $S$
and $\Omega_{\lambda}$ are obtained from the expression
(\ref{1.25}) by the replacement
$\Delta\rightarrow\Delta_{\textbf{q},\textbf{a}}$, $E\rightarrow
E_{\textbf{q},\textbf{a}}$,
$\varepsilon\rightarrow\varepsilon_{\textbf{q},\textbf{a}}$,
$f\rightarrow f_{\textbf{q},\textbf{a}}$. Hence the free energy of
a superconductor is
\begin{eqnarray}\label{3.15}
&&\Omega=-\frac{2i}{\beta}\lim_{\tau\rightarrow0^{-}}\sum_{\textbf{q}}\sum_{\textbf{k}}\sum_{n=-\infty}^{n=+\infty}
G\left(\textbf{k}+\frac{\textbf{q}}{2}-\frac{e}{2c}(\textbf{a}_\textbf{q}+\textbf{a}_\textbf{q}^{+},\omega_{n}\right)e^{-i\omega_{n}
\tau}\varepsilon\left(\textbf{k}+\frac{\textbf{q}}{2}-\frac{e}{2c}(\textbf{a}_\textbf{q}+\textbf{a}_\textbf{q}^{+})\right)\nonumber\\
&&+\sum_{\textbf{q}}\sum_{\textbf{k}}\frac{2}{\beta}\left[f_{\textbf{q},\textbf{a}}^{S}\ln
f_{\textbf{q},\textbf{a}}^{S}+(1-f_{\textbf{q},\textbf{a}}^{S})\ln
(1-f_{\textbf{q},\textbf{a}}^{S})-f_{\textbf{q},\textbf{a}}^{0}\ln
f_{\textbf{q},\textbf{a}}^{0}-(1-f_{\textbf{q},\textbf{a}}^{0})\ln
(1-f_{\textbf{q},\textbf{a}}^{0})\right]\nonumber\\
&&-\sum_{\textbf{q}}\frac{\lambda(\textbf{q})}{V}\sum_{\textbf{k}}w_{k}\frac{1}{\beta}\sum_{n=-\infty}^{n=+\infty}F_{\textbf{q},\textbf{a}}(\textbf{k},\omega_{n})
\sum_{\textbf{k}}w_{k}\frac{1}{\beta}\sum_{n=-\infty}^{n=+\infty}F_{\textbf{q},\textbf{a}}(\textbf{k},\omega_{n})\nonumber\\
&&+\frac{V}{8\pi}\sum_{\textbf{q}}\left(q^{2}(\textbf{a}_\textbf{q}\cdot\textbf{a}^{+}_\textbf{q})
    -(\textbf{q}\cdot\textbf{a}_\textbf{q})(\textbf{q}\cdot\textbf{a}^{+}_\textbf{q})\right)
\end{eqnarray}
where
\begin{equation}\label{3.16}
    f_{\textbf{q},\textbf{a}}^{S}(k)=\frac{1}{e^{\beta
E_{\textbf{q},\textbf{a}}(k)}+1},\qquad
f_{\textbf{q},\textbf{a}}^{0}(k)=\frac{1}{e^{\beta
|\varepsilon_{\textbf{q},\textbf{a}}\left(k\right)|}+1}
\end{equation}
are occupation numbers of states by boholons, and with help of the
term $f_{\textbf{q},\textbf{a}}^{0}$ the normal part of entropy is
separated, such that $\Omega_{S}(\Delta=0)=0$.

\emph{Unlike Ginzburg-Landau functional the obtained functional of
free energy} (\ref{3.15}) \emph{is correct for an arbitrary value
of the relation $\Delta(T)/T$, for arbitrary spatial variations of
the order parameter $\Delta(\textbf{r})$, for an arbitrary value
of a magnetic penetration depth $\Lambda(T)$ in comparison with a
coherent length $l_{0}$}. The observed value of
$\Delta_{\textbf{q},\textbf{a}}$ minimizes the free energy:
\begin{eqnarray}\label{3.15a}
\frac{d\Omega}{d\Delta_{\textbf{q},\textbf{a}}}=0&&\Longrightarrow(-i)\Delta_{\textbf{q},\textbf{a}}
=\frac{\lambda(\textbf{q})}{V\beta}\sum_{\textbf{k}}\sum_{n=-\infty}^{n=+\infty}
w_{k}F_{\textbf{q},\textbf{a}}(\textbf{k},\omega_{n})
\end{eqnarray}
If we suppose $\textbf{a}=0$ and neglect by the dependence on
$\textbf{q}$, then we obtain the expression (\ref{1.26}). The
functional (\ref{3.15}) is complicated for an analyze. Let us
suppose that $\Delta(\textbf{r})$ changes in space slowly and let
us expand the expression (\ref{3.15}) in degrees of
$\frac{\textbf{q}}{2}-\frac{e}{2c}(\textbf{a}_\textbf{q}+\textbf{a}_\textbf{q}^{+})$
keeping terms which are proportional to the vector in second
degree only. In order to simplify our consideration let us suppose
that the parameter of interaction
$\lambda(\textbf{q})=\texttt{const}$. This means that a spatially
inhomogeneity of the order parameter is caused by a magnetic field
only. Let us consider the high-temperature limit of free energy:
$\Delta\beta_{C}\ll 1$ at $T\rightarrow T_{C}$. This means, that
the expression (\ref{3.15}) can be expended in degrees of
$\Delta_{\textbf{q},\textbf{a}}$:
\begin{eqnarray}\label{3.19}
\Omega=\Omega_{n}&+&V\sum_{\textbf{q}}\left(\alpha(T)\Delta^{2}_{\textbf{q},\textbf{a}}
+\frac{1}{2}b\Delta^{4}_{\textbf{q},\textbf{a}}
+\gamma\left(\textbf{q}-\frac{e}{c}(\textbf{a}_\textbf{q}+\textbf{a}_\textbf{q}^{+})\right)^{2}\Delta_{\textbf{q},\textbf{a}}^{2}\right)\nonumber\\
&+&\frac{V}{8\pi}\sum_{\textbf{q}}\left(q^{2}(\textbf{a}_\textbf{q}\cdot\textbf{a}^{+}_\textbf{q})
    -(\textbf{q}\cdot\textbf{a}_\textbf{q})(\textbf{q}\cdot\textbf{a}^{+}_\textbf{q})\right),
\end{eqnarray}
where the coefficients $\alpha(T),b$ are determined by the formula
(\ref{1.33}), and the coefficient $\gamma$ is determined by the
formula (\ref{2.13}). The expansion (\ref{3.19}) has the form of
Ginzburg-Landau expansion of free energy in degrees of the order
parameter. Observed configurations of the order parameter
$\Delta_{\textbf{q},\textbf{a}}$ and magnetic field
$\textbf{a}(\textbf{q})$ minimizes the free energy:
\begin{eqnarray}
\frac{\delta\Omega}{\delta\Delta}=0 &\Rightarrow&
\alpha(T)\Delta_{\textbf{q},\textbf{a}}+b\Delta^{3}_{\textbf{q},\textbf{a}}
+\gamma\left(\textbf{q}-\frac{e}{c}(\textbf{a}_\textbf{q}+\textbf{a}_\textbf{q}^{+})\right)^{2}\Delta_{\textbf{q},\textbf{a}}=0 \label{3.20a}\\
\frac{\delta\Omega}{\delta\textbf{a}}=0 &\Rightarrow&
\textbf{j}^{+}(\textbf{q})=e\gamma\textbf{q}\Delta^{2}_{\textbf{q},\textbf{a}}
-\gamma\frac{e^{2}}{c}\Delta^{2}_{\textbf{q},\textbf{a}}(\textbf{a}_\textbf{q}+\textbf{a}_\textbf{q}^{+})\label{3.20b}\\
\frac{\delta\Omega}{\delta\textbf{a}^{+}}=0 &\Rightarrow&
\textbf{j}(\textbf{q})=e\gamma\textbf{q}\Delta^{2}_{\textbf{q},\textbf{a}}
-\gamma\frac{e^{2}}{c}\Delta^{2}_{\textbf{q},\textbf{a}}(\textbf{a}_\textbf{q}+\textbf{a}_\textbf{q}^{+})
\label{3.20c}
\end{eqnarray}
where $\textbf{j}(\textbf{q})$ is Fourier component of a current:
\begin{eqnarray}\label{3.22}
\textbf{J}(\textbf{r})=\frac{c}{4\pi}\texttt{rot}\textbf{H}(\textbf{r})&\Rightarrow&\textbf{j}(\textbf{q})
=-\frac{c}{4\pi}\textbf{q}\times\textbf{q}\times\textbf{a}_{\textbf{q}}
=-\frac{c}{4\pi}\left(\textbf{q}(\textbf{q}\cdot\textbf{a}_{\textbf{q}})-\textbf{a}_{\textbf{q}}q^{2}\right)\nonumber\\
&\Rightarrow&\textbf{j}^{+}(\textbf{q})=-\frac{c}{4\pi}\textbf{q}\times\textbf{q}\times\textbf{a}_{\textbf{q}}^{+}
=-\frac{c}{4\pi}\left(\textbf{q}(\textbf{q}\cdot\textbf{a}^{+}_{\textbf{q}})-\textbf{a}^{+}_{\textbf{q}}q^{2}\right).
\end{eqnarray}
From the equations (\ref{3.20b},\ref{3.20c}) we can see that
$\textbf{j}(\textbf{q})=\textbf{j}^{+}(\textbf{q})$. From the
equation (\ref{3.22}) we can see that
$\textbf{a}_{\textbf{q}}=\textbf{a}^{+}_{\textbf{q}}$.

If the superconductor is simply connected (without holes,
vortexes) we can pass to the transverse gauge of magnetic field:
$\textbf{q}\cdot \textbf{a}_{\textbf{q}}=0$. Then the current is
$\textbf{j}(\textbf{q})=\frac{c}{4\pi}\textbf{a}(\textbf{q})\textbf{q}^{2}$.
In transverse gauge the functional of free energy has the form:
\begin{equation}\label{3.23}
\Omega=\Omega_{n}+V\sum_{\textbf{q}}\left(\alpha(T)\Delta^{2}_{\textbf{q},\textbf{a}}+\frac{1}{2}b\Delta^{4}_{\textbf{q},\textbf{a}}
+\gamma\left(\textbf{q}^{2}+\frac{4e^{2}}{c^{2}}\textbf{a}_{\textbf{q}}^{2}\right)\Delta_{\textbf{q},\textbf{a}}^{2}\right)
+\frac{V}{8\pi}\sum_{\textbf{q}}\textbf{q}^{2}\textbf{a}_{\textbf{q}}^{2}.
\end{equation}
The equations of extremals are
\begin{eqnarray}
&&\alpha(T)\Delta_{\textbf{q},\textbf{a}}+b\Delta^{3}_{\textbf{q},\textbf{a}}
+\gamma\left(\textbf{q}^{2}+\frac{4e^{2}}{c^{2}}\textbf{a}_{\textbf{q}}^{2}\right)\Delta_{\textbf{q},\textbf{a}}=0 \label{3.24}\\
&&\textbf{j}(\textbf{q})=-8\gamma\frac{e^{2}}{c}\Delta^{2}_{\textbf{q},\textbf{a}}\textbf{a}_{\textbf{q}}\label{3.25}.
\end{eqnarray}
The closed currents $\textbf{q}\cdot\textbf{j}(\textbf{q})=0$
screen a magnetic field in a superconductor. The currents is
analogy to molecular currents of Ampere. From  Eq.(\ref{3.25}) one
can see, that the value
$Q=-8\gamma\frac{e^{2}}{c}\Delta^{2}_{\textbf{q},\textbf{a}}$ is
the kernel of magnetic response. The order parameter is function
of $\textbf{q}$ and $\textbf{a}(\textbf{q})$:
\begin{equation}\label{3.26}
\Delta^{2}_{\textbf{q},\textbf{a}}(T)=\frac{|\alpha(T)|}{b}\left(1-\frac{\gamma}{|\alpha(T)|}\left(\textbf{q}^{2}+\frac{4e^{2}}{c^{2}}\textbf{a}_{\textbf{q}}^{2}\right)\right)
\approx\frac{\Delta^{2}(T)}{\left(1+l^{2}(T)\left(\textbf{q}^{2}+\frac{4e^{2}}{c^{2}}\textbf{a}_{\textbf{q}}^{2}\right)\right)},
\end{equation}
where $ql(T)\ll 1$. From the formula (\ref{3.26}) one can see,
that the kernel $Q$ is a function of magnetic field. Hence the
electrodynamics of a superconductor is nonlinear. The kernel
considers both own nonlocality of a superconductor and the
nonlocality caused by fluctuations in the critical region (because
$l(T\rightarrow T_{C})\rightarrow\infty$). If to suppose
$\Delta=\texttt{const}$ at given temperature, then we shall obtain
London equation:
\begin{equation}\label{3.27}
\textbf{j}(\textbf{q})=-8\gamma\frac{e^{2}}{c}\Delta^{2}(T)\textbf{a}_{\textbf{q}}\equiv-\frac{c}{4\pi\Lambda^{2}(T)}\textbf{a}(\textbf{q})
\Rightarrow\Lambda^{2}(T)=\frac{c^{2}}{32\pi
e^{2}}\frac{b}{|\alpha(T)|\gamma},
\end{equation}
where $\Lambda(T)$ is the magnetic penetration depth in a
superconductor.

For research of the nonlocal characteristics of the free energy
functional (\ref{3.15}) let us consider the low-temperature limit
$\Delta\beta\gg 1$ at $T\rightarrow 0$. The value of gap is close
to its value at zero temperature $\Delta(T)\leq\Delta_{0}$.
Moreover, a magnetic field is weak, such that it changes the gap
lightly, that is the magnetic field is much smaller than critical
field $H\ll H_{C}$. Either as above, we assume that a change of
gap in space is slow. Starting from aforesaid and using the
expansion (\ref{1.29}) we obtain the free energy:
\begin{equation}\label{3.29}
\Omega=\Omega_{n}+V\sum_{\textbf{q}}\left(\alpha_{0}(T)+b_{0}(T)\Delta_{\textbf{q},\textbf{a}}+d_{0}\Delta^{2}_{\textbf{q},\textbf{a}}
+\gamma\left(\textbf{q}^{2}+\frac{4e^{2}}{c^{2}}\textbf{a}_{\textbf{q}}^{2}\right)\Delta_{\textbf{q},\textbf{a}}^{2}\right)
+\frac{V}{8\pi}\sum_{\textbf{q}}q^{2}a_{\textbf{q}}^{2},
\end{equation}
where coefficients $\alpha_{0}(T),b_{0}(T),d_{0}$ are determined
by the formulas (\ref{1.30}), and coefficient $\gamma$ is
determined by the formula (\ref{2.13}), the magnetic field is
considered in the transverse gauge $\textbf{q}\cdot
\textbf{a}_{\textbf{q}}=0$. The observed configurations of order
parameter $\Delta_{\textbf{q},\textbf{a}}$ and magnetic field
$\textbf{a}(\textbf{q})$ minimized the free energy:
\begin{eqnarray}
  \frac{\delta\Omega}{\delta\Delta}=0 &\Rightarrow& b_{0}(T)+2d_{0}\Delta_{\textbf{q},\textbf{a}}
  +\gamma\left(\textbf{q}^{2}+\frac{4e^{2}}{c^{2}}\textbf{a}_{\textbf{q}}^{2}\right)\Delta_{\textbf{q},\textbf{a}}=0 \label{3.30}\\
  \frac{\delta\Omega}{\delta\textbf{a}}=0 &\Rightarrow&
\textbf{j}(\textbf{q})=-8\gamma\frac{e^{2}}{c}\Delta^{2}_{\textbf{q},\textbf{a}}\textbf{a}(\textbf{q})\label{3.31}.
\end{eqnarray}
If in the equation (\ref{3.31}) to assume $\Delta=\texttt{const}$
at given temperature, then we shall have London equation again:
\begin{equation}\label{3.35}
\textbf{j}(\textbf{q})=-8\gamma\frac{e^{2}}{c}\Delta^{2}(T)\textbf{a}(\textbf{q})\equiv-\frac{c}{4\pi\Lambda^{2}(T)}\textbf{a}(\textbf{q})
\Rightarrow\Lambda^{2}(T)=\frac{c^{2}}{8\pi
e^{2}}\frac{d_{0}^{2}}{b_{0}^{2}(T)\gamma}.
\end{equation}
The equations (\ref{3.30},\ref{3.31}) allows to generalize London
equation. From the equation (\ref{3.30}) we can find the value of
gap:
\begin{equation}\label{3.36}
    \Delta_{\textbf{q},\textbf{a}}(T)=\frac{-b_{0}(T)}{2d_{0}+2\gamma\left(\textbf{q}^{2}+\frac{4e^{2}}{c^{2}}\textbf{a}_{q}^{2}\right)}
\end{equation}
Then the equation for current has the form:
\begin{equation}\label{3.37}
\textbf{j}(\textbf{q})=-8\gamma\frac{e^{2}}{c}\frac{b_{0}^{2}(T)}{\left(2d_{0}+2\gamma\left(\textbf{q}^{2}
+\frac{4e^{2}}{c^{2}}\textbf{a}_{\textbf{q}}^{2}\right)\right)^{2}}\textbf{a}(\textbf{q})
\equiv Q(\textbf{q},\textbf{a})\textbf{a}(\textbf{q})
\end{equation}
This equation is the nonlocal and nonlinear generalization of
London equation in the long wavelength limit $q\rightarrow 0$,
because the kernel $Q$ is the function of $\textbf{q}$ and
magnetic field $\textbf{a}(\textbf{q})$. However the equation
(\ref{3.37}) is correct when the magnetic field is much weaker
than the critical field $H\ll H_{C}$.

Let us neglect the nonlinearity, that is we suppose that the
kernel $Q$ is function of $\textbf{q}$ only. Then we have:
\begin{equation}\label{3.38}
\textbf{j}(\textbf{q})=-8\gamma\frac{e^{2}}{c}\frac{b_{0}^{2}(T)}{\left(2d_{0}+2\gamma\textbf{q}^{2}\right)^{2}}\textbf{a}(\textbf{q})
=-8\gamma\frac{e^{2}}{c}\frac{\Delta_{0}^{2}(T)}{\left(1+\frac{\gamma}{d_{0}}\textbf{q}^{2}\right)^{2}}\textbf{a}(\textbf{q})
\approx-8\gamma\frac{e^{2}}{c}\frac{\Delta_{0}^{2}(T)}{\left(1+l_{0}^{2}\textbf{q}^{2}\right)}\textbf{a}(\textbf{q})
\equiv Q(q)\textbf{a}(\textbf{q}),
\end{equation}
where $l_{0}q\ll 1$,
$\frac{\gamma}{d_{0}}=\frac{l_{0}^{2}\nu_{F}}{\nu_{F}}=l^{2}_{0}$
is the coherent length at temperature $T=0$. Thus \emph{we have
the nonlocal kernel} $Q(q)$, \emph{where radius of nonlocality is
equal to the coherent length} $l_{0}$. \emph{This result
corresponds to nonlocal Pippard electrodynamics} (in the long wave
limit). \emph{This fact proves a nonlocality of the obtained free
energy functional} (\ref{3.15}). For generalization in the case of
large value $l_{0}q$ it is necessary to expand the free energy
(\ref{3.15}) in degrees of $\textbf{q}$.

\section{Conclusion.}\label{conclusion}

In this paper we developed the microscopic approach for a finding
of the free energy functional of a superconductor with help of
direct calculation of a vacuum amplitude. The functional is
calculated on the dressed propagators, which takes into account
the interaction of a free fermion with the fluctuations of pairing
(with the condensate of pairs). As a result of such interaction
the dispersion law of quasi-particles is changed and anomalous
propagators appear. This means, that a spontaneous symmetry
breakdown takes place. After consideration of the interaction of
particles with the fluctuations of pairing all characteristics of
a system must be calculated over the new vacuum with the broken
symmetry. As a result, the free energy is function of the
amplitudes $\Omega=\Omega(\Delta\Delta^{+})$, and their observed
values minimize the free energy.

With help of the developed microscopic approach we have obtained
the free energy of a spatially inhomogeneous superconductor (the
system with a spatially inhomogeneous distribution of the order
parameter $\Delta(\textbf{r})$) for all temperatures. The cause of
the inhomogeneity is the dependence of the interaction constant of
electrons on coordinates and the influence of a magnetic field.
The obtained expression for the free energy is valid for arbitrary
spatial variations of the order parameter. In the long-wave limit
the functional is similar to Ginzburg-Landau expansion, but the
coefficients of expansion depend on coordinates. Moreover, the
obtained expression shows the nonlocality of the order parameter:
the value of energy gap in a point is determined by the
distribution of the energy gap in some neighborhood of this point.

Above-mentioned results have been generalized on the case of
presence of a magnetic field and a current. The analysis of the
obtained free energy functional (\ref{3.15}) shows, that the
nonlocality of a magnetic response of a superconductor is the
result of the nonlocality of order parameter. The equations of
superconductor's state are extremals of the functional, and they
are obtained by variation over the gap $\Delta$ and the magnetic
field $\textbf{a}$. In the high-temperature limit the obtained
equations are analogous to Ginzburg-Landau equations. In the
low-temperature limit the equations show the nonlocal connection
between the magnetic field and the current according to the
Pippard low.


\begin{thebibliography}{99}



\bibitem{gog1} G. A. Gogadze, A. N. Omel`yanchuk, Fizica nizkikh temperatur \textbf{22}, No.6, p. 648-651 (1996)

\bibitem{gog2} Gogadze G. A., Fizica Nizkikh Temperatur \textbf{21}, No.2, p. 177-182 (1995)


\bibitem{kapaev} Kapaev A. V., Kopaev Yu. V., JETP Letters  \textbf{68}, No.3, p. 211-216 (1998)

\bibitem{asker} I.N. Askerzade, Physics-Uspekhi \textbf{52}, p. 977-988 (2009)

\bibitem{lawr} W.E. Lawrence, S. Doniach, in Proc. of the 12th Intern. Conf. on
Low Temperature Physics, Kyoto, 1970 (Ed. E Kanda) (Tokyo:
Kcegaku, 1971) p. 361

\bibitem{ovch1} Ovchinnikov Y. N., Kresin V. Z. Eur. Phys. J. B \textbf{45}, No.1 p.5-7 (2005)

\bibitem{ovch2} Ovchinnikov Y. N., Kresin V. Z. Eur. Phys. J. B \textbf{47}, No.3 p.333-336 (2005)

\bibitem{ovch3} Kresin V. Z., Ovchinnikov Y. N. Phys. Rev. B \textbf{81}, No.21 p.214505 (2010)

\bibitem{khlyus} I.N. Khlyustikov, A.I. Buzdin, Physics-Uspekhi \textbf{31}, p. 409–433 (1988)


\bibitem{ivan1} Yu.M. Ivanchenko, T.K. Soboleva, JETP letters. \textbf{51}, No.2, p.114-117 (1990).

\bibitem{ivan2} Yu.M. Ivanchenko, T.K. Soboleva. Phys. Lett. \textbf{A147}, No.1, p.65-69 (1990).

\bibitem{mints}R.G. Mints, I.B. Snapiro. Phys. Rev. \textbf{B51}, No.5, p.3054-3057 (1995).

\bibitem{gurev} A. Gurevich. Phys. Rev. \textbf{B46}, No.5, p.3187-3190 (1992).

\bibitem{silin1} V.P. Silin., JETP letters. \textbf{58}, No.9, p.701-704 (1993).

\bibitem{silin2} V.P. Silin., JETP letters. \textbf{60}, No.6, p.460-463 (1994).

\bibitem{degen} P.G. Gennes, \emph{Superconductivity of metall and alloys}
(W. A. Benjamin, inc., New York - Amsterdam, 1968).

\bibitem{tew} L. Teword, Phys. Rev. E \textbf{132}, No2, p.595 (1963).

\bibitem{werth} N.R. Werthamer, Phys. Rev. E \textbf{132}, No2, p.663 (1963).

\bibitem{hook} J.R. Hook, J.R. Waldram, Proc. R. Soc. Lond. A. \textbf{334} No.1597, p.171-192 (1973)

\bibitem{matt1} Richard D. Mattuk, \emph{A guide to feynman diagrams in the many-body problem}
(H. C. Oersted Institute University of Copenhagen, Denmark, 1967).

\bibitem{matt2} R. D. Mattuk, B.Johansson, Advances in Physics \textbf{17},
p.509 (1968).

\bibitem{sad} M.V. Sadovskii, \emph{Superconductivity and Localization}
(World Scientific, Singapore, 2000).

\bibitem{migdal}  A. B. Migdal, \emph{Theory of Finite Fermi Systems and Applications
to Atomic Nuclei} (Interscience Publishers,  New York, 1967)

\bibitem{pines}  D. Pines, \emph{The many-body problem} (University of Illinois,  New York, 1961)

\end{thebibliography}
\end{document}